\documentclass[]{aastex631}

  \usepackage{amsmath}
  \usepackage{graphicx}
  \usepackage[utf8]{inputenc}
\renewcommand{\arraystretch}{1.5}
  \graphicspath{{./}{figures/}}

  \received{--}
  \revised{--}
  \accepted{--}
  \shorttitle{Solar differential rotation}
\begin{document}

\title{Investigating Variations in Solar Differential Rotation by Helioseismology}

   \author[0000-0003-3067-288X]{Krishnendu Mandal}
  \affiliation{New Jersey Institute of Technology, Newark, NJ 07102, USA}
  
  \author[0000-0003-0364-4883]{Alexander G. Kosovichev}
  \correspondingauthor{Krishnendu Mandal}
  \email{krishnendu.mandal@njit.edu}
  \affiliation{New Jersey Institute of Technology, Newark, NJ 07102, USA}
  \affiliation{NASA Ames Research Center, Moffett Field, CA 94035, USA}
  \author[0000-0001-9884-1147]{Valery V. Pipin}
  \affiliation{Institute of Solar-Terrestrial Physics, Russian Academy of Sciences, Irkutsk, 664033, Russia}
  \author[0000-0003-1531-1541]{Sylvain G. Korzennik}
  \affiliation{Center for Astrophysics, Harvard \& Smithsonian, Cambridge, MA 02138, USA}
\begin{abstract}
 Helioseismic signatures of dynamo waves have recently been discovered in variations of the solar differential rotation, offering valuable insights into the type of dynamo mechanism operating in the solar convection zone. To characterize these variations, we analyze p-mode frequency-splitting data estimated using time intervals of various lengths to enhance the signal-to-noise ratio in inversions of zonal flows. We introduce a novel time-dependent inversion method that inherently smooths the solution over time, eliminating the need for separate post-processing smoothing. By applying this approach to observational data from the SOHO Michelson Doppler Imager, SDO Helioseismic Magnetic Imager, and Global Oscillation Network Group, we identify similar dynamo wave patterns in both the zonal acceleration and the zonal flow throughout the entire convection zone. Our analysis shows that while using longer time series smooths out temporal variations, the fundamental features observed in the short time series (i.e.\ 72-day long) persist when inverting datasets covering different time periods.  These findings reinforce earlier detections and offer further validation of solar dynamo models. We additionally investigate the dimensionless radial gradient of rotation. Its value is close to $-1$ and increases in the deeper layers, remaining nearly constant from the equator to mid-latitudes within the depth range of 13 to 35 Mm below the surface; the results at high latitudes remain somewhat inconclusive. The variation of this quantity displays a torsional oscillation–like pattern, albeit with certain differences.
 
\end{abstract}
  
 \section{Introduction} \label{sec:intro}
Torsional oscillations (zonal flows migrating from high to low latitudes during solar cycles) were first identified on the solar surface in Doppler velocity maps by \citet{howard80}. After removing the mean differential rotation, the data revealed alternating bands of fast and slow zonal flows, originating at mid-latitudes and migrating toward the equator, resembling the magnetic butterfly diagram. These zonal flows, termed torsional oscillations due to their cyclic nature, were initially suggested to be driven by the magnetic fields of active regions. Numerical 3D simulations by \citet{gustavo16} indicated that torsional oscillations in the model arise from magnetic torque at the base of the convection zone. Additionally, \citet{pipin2019} showed that the modulation of turbulent heat transport by magnetic fields within the convection zone induces variations in zonal and meridional flows, leading to an extended 22-year cycle of torsional oscillations. Several dynamo models have been proposed to explain the 11-year periodic variations of the solar cycle \citep[e.g., ][]{parker_55_dynamo,babcock61,arnab_dikpati99,dikpati09,paul11}. Since the magnetic field within the solar interior cannot be directly observed, the only way to validate these mechanisms is by analyzing how the solar internal flows evolve throughout the solar cycle.

Helioseismology provides a powerful tool for probing the subsurface flow profile of solar zonal flows \citep{howe18,basu19}. A comprehensive review of global helioseismology can be found in \citet{basu_2016}. Using observations of rotational splitting of solar oscillation frequencies \citep[][]{larson2015,larson18,korzennik2013} derived from observations taken by the  Michelson Doppler Imager \citep[MDI;][]{scherrer95} onboard the Solar and Heliospheric Observatory (SOHO), \citet{sasha97} discovered that torsional oscillations extend beneath the solar surface, while \citet{vorontsov02} later demonstrated that these oscillations span the entire convection zone, with their phase propagating both poleward and equatorward from mid-latitudes at all depths throughout the convective envelope. 
Recently, dynamo-wave-like signatures have been detected in variations of solar differential rotation, commonly known as torsional oscillations \citep[][]{sasha2019,mandal24_dw}. \citet{mandal24_dw} observed similar oscillations when inverting rotational frequency splittings derived using the Global Oscillations Network Group \citep[GONG;][]{hill_1996_GONG} observations, further supporting the earlier detection obtained using observations acquired with the Helioseismic and Magnetic Imager \citep[HMI;][]{hmi,scherrer2012}, on board the Solar Dynamics Observatory (SDO)  by \citet{sasha2019}. 
The rotational frequency splittings derived using  time series of different lengths and all three instruments, and with the same data analysis methodology \citep{sylvain23}, motivated us to investigate these findings using these multiple datasets and different time intervals. 

Since our focus is on time variations, we must ensure both high temporal resolution and good signal-to-noise ratio from the surface down to the base of the convection zone. This requires a careful trade-off between the length of the analyzed epoch and the corresponding time resolution. In our previous work, \citep{mandal24_dw}, we found that dynamo waves originate in the deeper convection zone and propagate to the surface rapidly, within a year at high latitudes, but take approximately 5–6 years to reach mid and low latitudes. In this study, we quantify these values, as they provide crucial constraints for dynamo models. A successful dynamo model must account for these aspects of torsional oscillations. Since both convective flows and magnetic fields influence the solar dynamo through Lorentz force feedback -- where changes in flow affect the magnetic field and vice versa -- a comprehensive model should also be able to explain the observed variations in solar differential rotation. 

We use frequency splitting data obtained using different lengths of time series as provided by \citet{sylvain23}. The goal is to achieve a high signal-to-noise ratio for deducing properties of the evolving zonal flows. As shown in Figure 5 of our previous work \citep{mandal24_dw}, using a longer time series of data reduces noise and, therefore, can be used to infer changes with the solar cycle with better accuracy. Typically, the inversion is performed for each time segment, followed by temporal smoothing, usually using a Gaussian filter. In this work, we propose a novel  technique that operates jointly in time and radius, thereby eliminating the need for temporal smoothing. Additionally, we introduce an alternative smoothing approach, apart from Gaussian smoothing, for cases where the inversion is performed separately for each time segment.    

Determining the radial gradient of rotation in the near-surface shear layer (NSSL) has recently become an important research focus due to its key role in the solar dynamo \citep{brandenburg_2005, pipin2011}. 
\citet{corbard2002} studied the radial gradient of the rotation rate using MDI surface gravity mode observations and found that the logarithmic gradient of the rotation frequency with respect to radius is approximately $-1$. They also reported that this gradient decreases in absolute magnitude above $30^\circ$ latitude. This result was later revisited by \citet{barekat2014}, who found that the gradient remains constant up to \(60^\circ\) latitude.
More recently, \citet{antia2022,komm2023,rozelot2025} estimated the radial gradient of the rotation rate and reported that its value varies with depth, starting with a higher absolute magnitude near the surface and decreasing significantly toward the base of the near-surface shear layer. They also observed that variations in the radial gradient of the rotation rate, defined as a deviation from the radial gradient averaged over time, $\delta\nabla_{r}(\Omega)$, exhibit torsional oscillation-like behavior. Namely,
\begin{equation}
    \delta(\nabla_{r}(\Omega)) = \nabla_r(\Omega) - \langle \nabla_r(\Omega) \rangle_t 
\end{equation}
and
\begin{equation}
    \nabla_ r(\Omega) = \frac{\partial\log\Omega}{\partial\log r}
\end{equation}

In this work, we further investigate these findings using our analysis. Ongoing research is increasingly exploring the role of NSSL in solar dynamo processes, with several studies investigating its potential impact \citep[e.g.][and references therein]{karak2016,irina2023}. \citet{Dikpati2002} studied the effect of NSSL on the flux transport dynamo. \citet{brandenburg_2005} argued that the NSSL may play a critical role in the distributed dynamo, and according to the theory of \citet{parker_55_dynamo}, the radial shear revealed by helioseismology might explain the equatorward migration of the solar activity belt. The formation of the NSSL still remains an active research topic \citep[][]{jha_2021}, particularly regarding its significance for the solar dynamo \citep{pipin2011}. Given the important role the NSSL plays in solar dynamics, it is essential to characterize the rotation in this layer.

 \section{Data Analysis}
  The rotational p-mode frequency data available for different time lengths \citep{sylvain23} allow us to assess the accuracy of our results. Specifically, we use results from fitting time series of $1 \times 72$-day, $4 \times 72$-day, $5 \times 72$-day and $8 \times 72$-day lengths using the HMI, MDI, and GONG data series and available at the Joint Science Operations Center (JSOC) under the following data series names: \texttt{su\_sylvain.hmi\_V\_sht\_modes\_asym\_v7}, \texttt{su\_sylvain.mdi\_V\_sht\_modes\_asym\_v7}, and \texttt{su\_sylvain.gong\_V\_sht\_modes\_asym\_v7} for HMI, MDI and  GONG observations respectively.
  From the JSOC of SDO, we use MDI data from \texttt{mdi.vw\_V\_sht\_modes} and HMI data from \texttt{hmi.V\_sht\_modes}. These JSOC pipeline data products are available in 72-day and $5 \times 72$-day series; in this work, we use the $5 \times 72$-day data products from JSOC. The GONG data series spans from May 7, 1995, to March 23, 2024; the MDI data series from May 1, 1996, to March 20, 2011, while the HMI data series spans from April 30, 2010, to August 14, 2024. There is a one-year overlap between MDI and HMI, and during this period, we use the HMI data. These time intervals are sufficiently long to capture cycles 23 and 24 and the first half of cycle 25 as it approaches its maximum. 
  
 The frequency of an acoustic mode is generally denoted as $\nu_{n\ell m}$, where $(n, \ell, m)$ represent the radial order, harmonic degree, and azimuthal order, respectively. In a spherically symmetric Sun, the frequencies are degenerate in $m$, however, solar rotation lifts this degeneracy, which can be parametrized as follows
  \begin{equation}
      \nu_{n\ell m}=\nu_{n\ell}+\sum_{s}^{s_{\text{max}}} a_{s}(n,\ell)\mathcal{P}_{s}^{\ell}(m),
      \label{eq:nu_lmn}
\end{equation}
where, $\nu_{n\ell}$ represents the central frequency of a multiplet, while $a_s$ are the rotational splitting coefficients, and $\mathcal{P}_{s}^{\ell}$ are polynomials of degree $s$ \citep[see][]{ritzwoller91}. The odd-order a-coefficients provide information about solar rotation, owing to the hemispherically symmetric nature of the oscillatory pattern. In contrast, the even-order a-coefficients are sensitive to structural asphericities, magnetic fields, and second-order rotational effects.

To validate our analysis, we conduct forward modeling and inversion using synthetic frequency splittings calculated for the dynamo model of \citet{pipin2019}. After successfully verifying our method, we apply it to the observed frequency-splitting data. In the next section, we provide a detailed discussion of the forward modeling approach.

 \subsection{Forward Modeling}
  The solar rotational velocity can be represented as follows 
   \begin{equation}
    V(r,\theta,t)=\sum_{s=1,3,5, \cdots} w_{s}(r,t)\frac{\partial Y_{s,0}(\theta,\phi)}{\partial\theta}
    \label{eq:v_w}
   \end{equation}
   where $Y_{s,0}(\theta,\phi)$ represents a spherical harmonic with angular degree $s$ and zero azimuthal order, $w_s(r)$ are the radial coefficients in the velocity expansion, $r$ denotes the radius from the solar center, $\theta$  is the co-latitude, and $\phi$ is the azimuthal angle.  The rotation of the Sun causes p- and f-mode frequencies to split as a function of the azimuthal order $m$ of the oscillation eigenfunctions. The relationship between the frequency splitting  and the radial velocity coefficients can be expressed through the following integral equation: 
   \begin{equation}
       a_{s}(n,\ell;t)=\int_{0}^{R} K_{s}(n,\ell;r)w_{s}(r,t)r^2 dr, 
       \label{eq:a_coeff}
   \end{equation}
   where $K_{s}(n,\ell)$ is the sensitivity kernel for the mode $(n,\ell),$  indicating how strongly this mode is affected by rotation at spatial scale $s$. In our previous work \citep{mandal24_dw}, we used generalized spherical harmonics to derive the sensitivity kernel, following a similar approach to that of \citet{ritzwoller91}. The use of generalized spherical harmonics allows for a more elegant and structured formulation, making it easier to handle complex algebra, as demonstrated in \citet{mandal24_dw}. For forward modeling, we use the varying with time rotation profile $V(r,\theta,t)$ from \citet{pipin20}, derived from their dynamo simulation. This profile is then decomposed into a spherical harmonic basis to obtain its component form $ w_s$. Finally, we substitute $w_s$ into Equation \ref{eq:a_coeff} to compute the frequency splitting coefficients $a_s$. The estimated $a_s$ values are then used for inversion to reconstruct the $w_s$ profile. The inversion method that we employ is discussed in the following section, while an alternative approach can be found in our previous work \citet{mandal24_dw}. The inversion, typically performed separately for each time segment as described in \citet{mandal24_dw}, is usually followed by temporal smoothing. This is commonly done using Gaussian smoothing over time. In Appendix \ref{sec:app_smooth}, we discuss how alternative smoothing approaches can be applied. 
  
 \subsection{Inversion Method}
  After obtaining the forward-modeled a-coefficients, we seek to determine whether we can recover the original profile $V(r,\theta)$ used to generate them. To simulate realistic conditions, we add random noise to the forward-modeled a-coefficients based on observational uncertainties. Using a Regularized Least Squares (RLS) approach, we minimize the misfit function, $\chi$, as follows
  \begin{equation}
      \chi^2=\sum_{n,\ell,t_{k}}\frac{(a_{s}(n,\ell;t_{k})-\int\mathcal{K}_{n,\ell}w_{s}(r,t_{k})r^{2}dr)^{2}}{\sigma_{s}(n,\ell;t_{k})^{2}}+\lambda_{r}\int \left( g(r)\frac{d^{2}w_{s}(r,t_{k})}{dr^{2}}\right)^{2}dr+\lambda_{t}\sum_{t_{k}}\left(f(r)\frac{d w_{s}(r,t)}{dt}\right)^{2}\vert_{t_{k}},
      \label{eq:chi}
  \end{equation}
  $f(r)$ and $g(r)$ are two functions used to smooth the solution in both time and radius. To achieve greater smoothing in deeper layers, an appropriate functional form should be chosen that increases as the radius decreases. This is the first time a rotational inversion has been performed simultaneously in both time and radius, treating time as an additional dimension. Previously, such inversions were typically carried out in radius and latitude, or in radius alone. Alternative inversion methods to the Regularized Least Squares  (RLS) include the Optimally Localized Averages \citep[OLA;][]{backus_1968_ola, backus_1970_OLA} and the Subtractive Optimally Localized Averages \citep[SOLA;][]{pijpers_1994_SOLA}. Interested readers may also refer to the different inversion techniques used in a similar study by \citet{schou98}. Recently, \citet{sylvain_2024_inv} proposed an iterative inversion method using the Simultaneous Algebraic Reconstruction Technique (SART), originally developed for image reconstruction. Exploring different avenues of analysis — from frequency measurements to inversion methods — is crucial for obtaining an accurate understanding of the solar rotation profile from the surface down to the deeper layers.  

 Inversion is an ill-posed problem, hence it is necessary to impose additional constraints. In the RLS method, we assume that the solution is smooth. In order to impose a smoothness constraint on the solution, we apply second-order derivative smoothing in radius and first-order derivative smoothing in time to $w_s(r,t)$. The function, $w_s$, is expanded on the basis of B-spline functions with $50$ equally spaced knots along the acoustic depth, ranging from the surface to the center, and a third temporal grid point as a knot in the time direction. Additional details of our inversion method are described in Appendix \ref{sec:app_inv}. We estimate error bars using a Monte Carlo method by adding noise based on observed noise to the estimated splitting coefficients, using a set of 100 realizations. Each set is inverted, and the standard deviation of these values gives us an estimate of the error in $w_s$. Finally, by substituting  $w_s$ into Equation \ref{eq:v_w} we obtain the error in velocity, $V(r,\theta,t)$. 
 \begin{figure}
     \centering
     \includegraphics[scale=0.42]{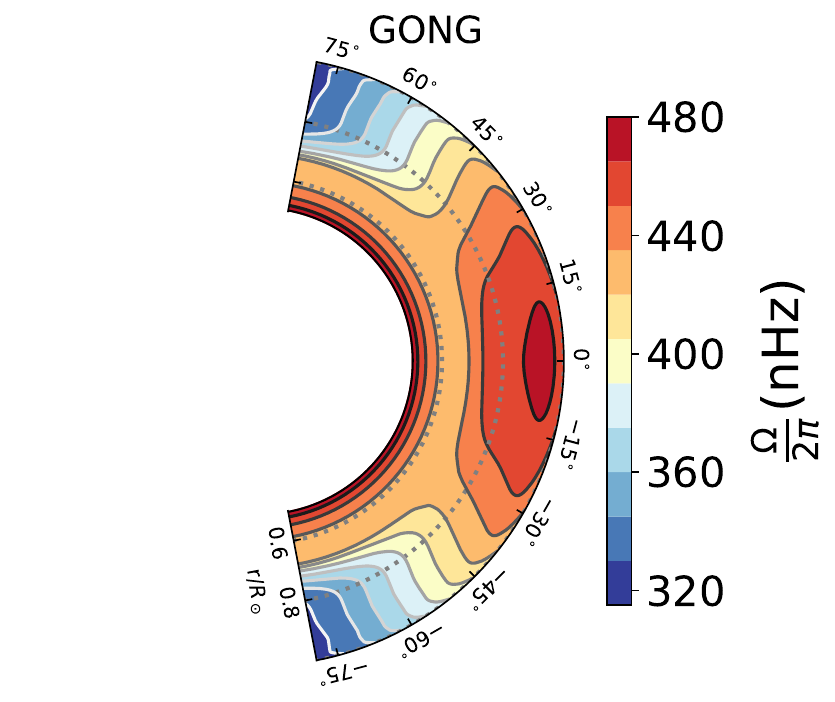}\includegraphics[scale=0.42]{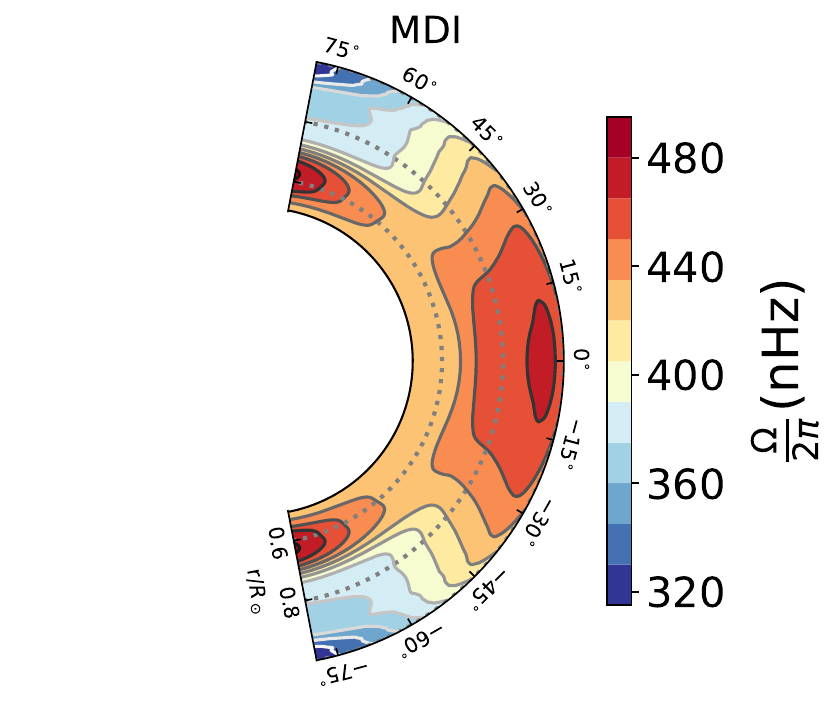}
    \includegraphics[scale=0.42]{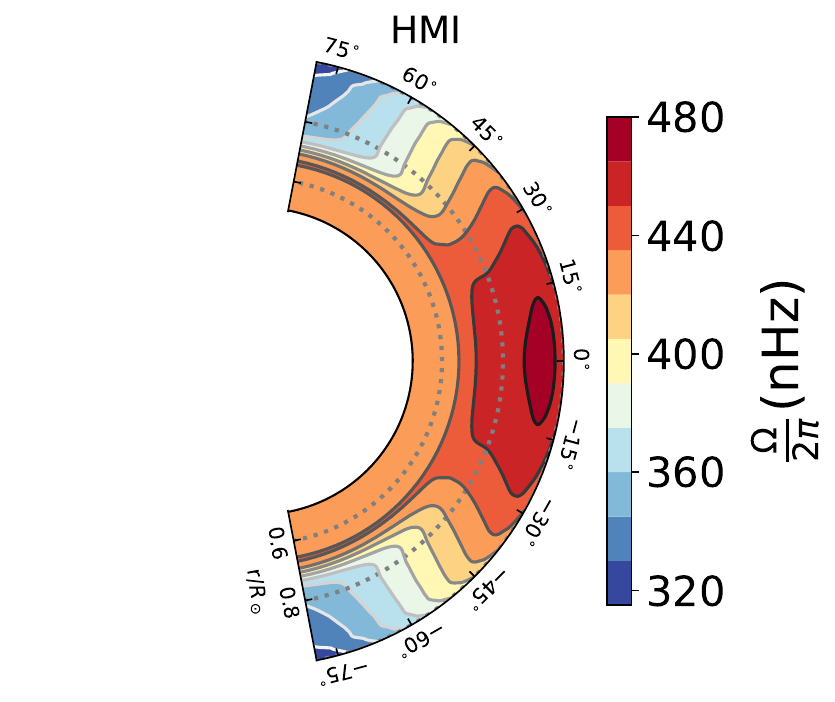}
     \caption{The mean differential rotation profile is shown from left to right using observations from GONG, MDI, and HMI, respectively. For all instruments, we use $5\times 72$-day data sets based on the data products of \citet{sylvain23}. We have used entire time periods of each instrument. High latitudes ($75^\circ$) and deeper regions in the radiative interior are subject to large uncertainties due to poorly localized averaging kernels. Therefore, we exclude those regions from the plot.}
     \label{fig:contour_fig}
 \end{figure}
   \begin{figure}
      \centering
      \includegraphics[scale=0.4]{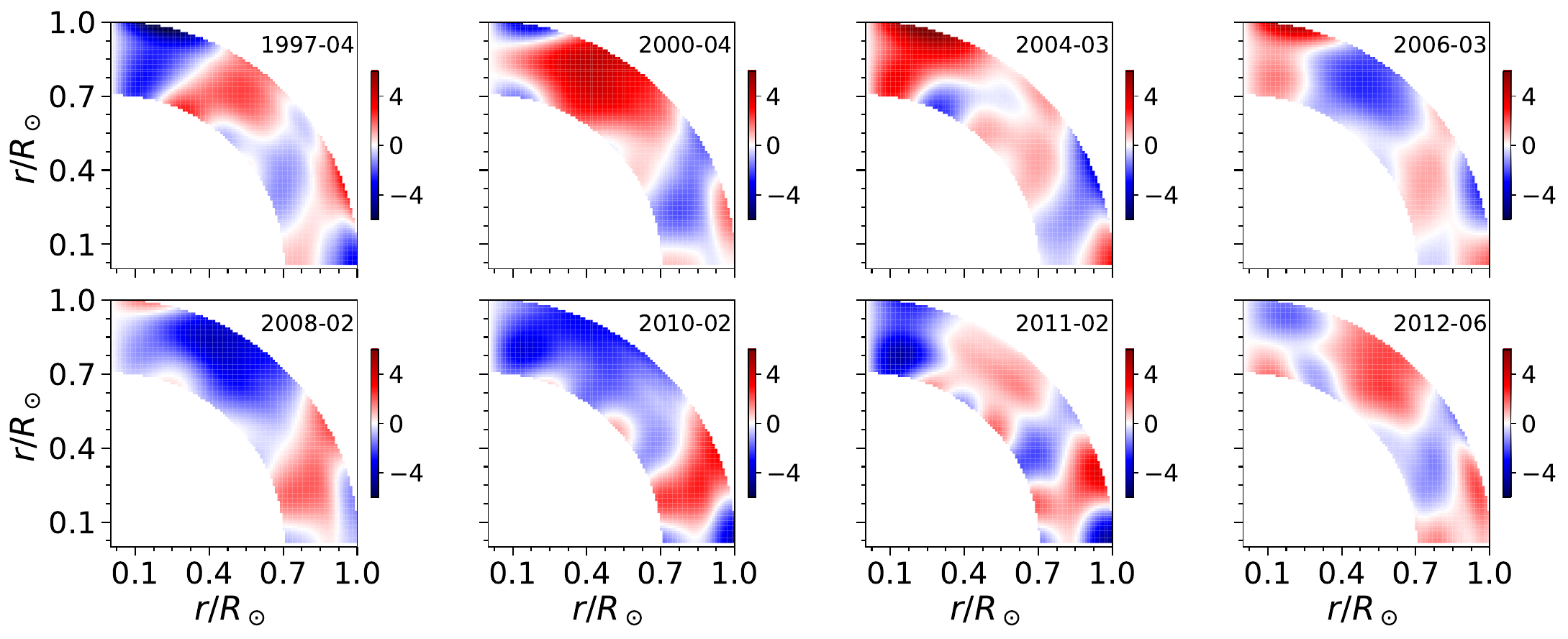}
      \caption{Evolution of solar zonal flow in the solar convection zone from the analysis of MDI and HMI combined data sets. We use $5\times 72$-days datasets from \citet{sylvain23} for this analysis. We plot the torsional oscillation velocity (measured in m/s) for different time (mentioned as well in each panel) to illustrate the evolution of zonal flow. The total time span is chosen to extend beyond the duration of Solar Cycle 23, enabling us to capture the flow evolution just before the cycle's onset and the emergence of the next cycle. It is observed that the low-latitude branch takes some time to migrate from high to low latitudes, while changes near high latitudes occur very quickly. This effect is especially noticeable at high latitudes in the panels for 2004-03 and 2006-03, and also in the transition between 2010-02 and 2011-02.}
      \label{fig:dynamo_still}
  \end{figure}
  
  \section{Results} 
 \subsection{Torsional Oscillations}
 In Figure \ref{fig:contour_fig}, we show contour plots for the differential rotation obtained from GONG, MDI, and HMI observations. For analysis of the MDI data,  we have used harmonic degree $\ell<120$ as suggested earlier by \citet{antia08,basu19}, which ensures results are consistent with other instruments. This is also evident in Figure 2 of our previous work, where we plot the temporal variation of the a-coefficients measured for modes whose ray turning points are located near the solar surface. For MDI, we need to exclude modes with harmonic degrees $>120$ to ensure that the mean a-coefficient variations from all three instruments are consistent with each other.   
 The solar differential rotation deviates from the Taylor-Proudman theorem, which predicts the cylinder-like angular velocity distributions of a rotating fluid. The Near-Surface Shear Layer (NSSL) extends up to approximately $75^\circ$ latitude. Below the NSSL, the solar rotation exhibits minimal radial variation. Another shear layer is present between $0.8\,R_\odot$  and $0.65\, R_\odot$, below which the rotation becomes uniform throughout the radiative interior.
 %that in a rotating fluid, perturbations from a solid body tend to produce columnar structures aligned with the rotation axis.
 
   \begin{figure}
      \centering
      \includegraphics[scale=0.4]{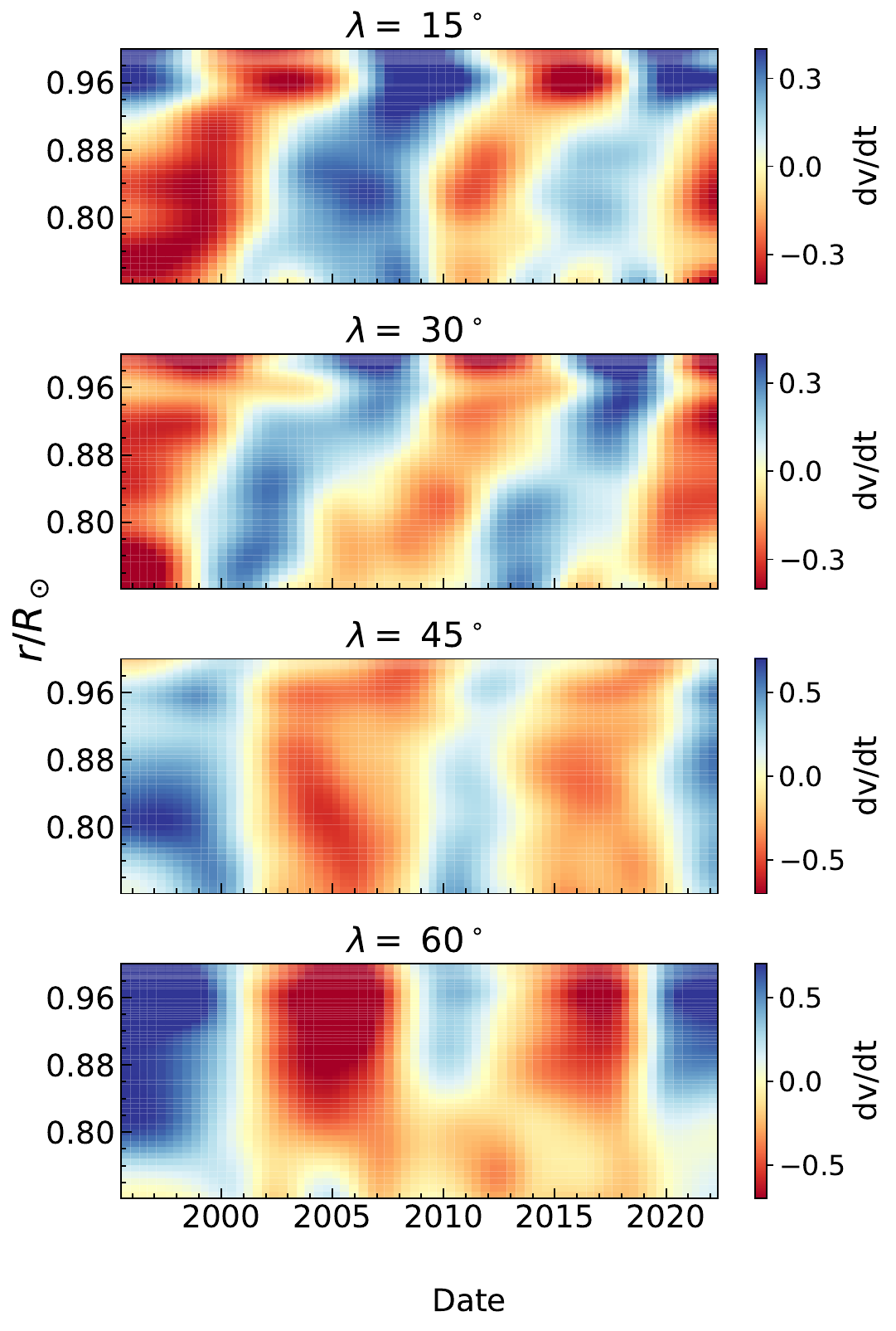}\hspace{1em}\includegraphics[scale=0.4]{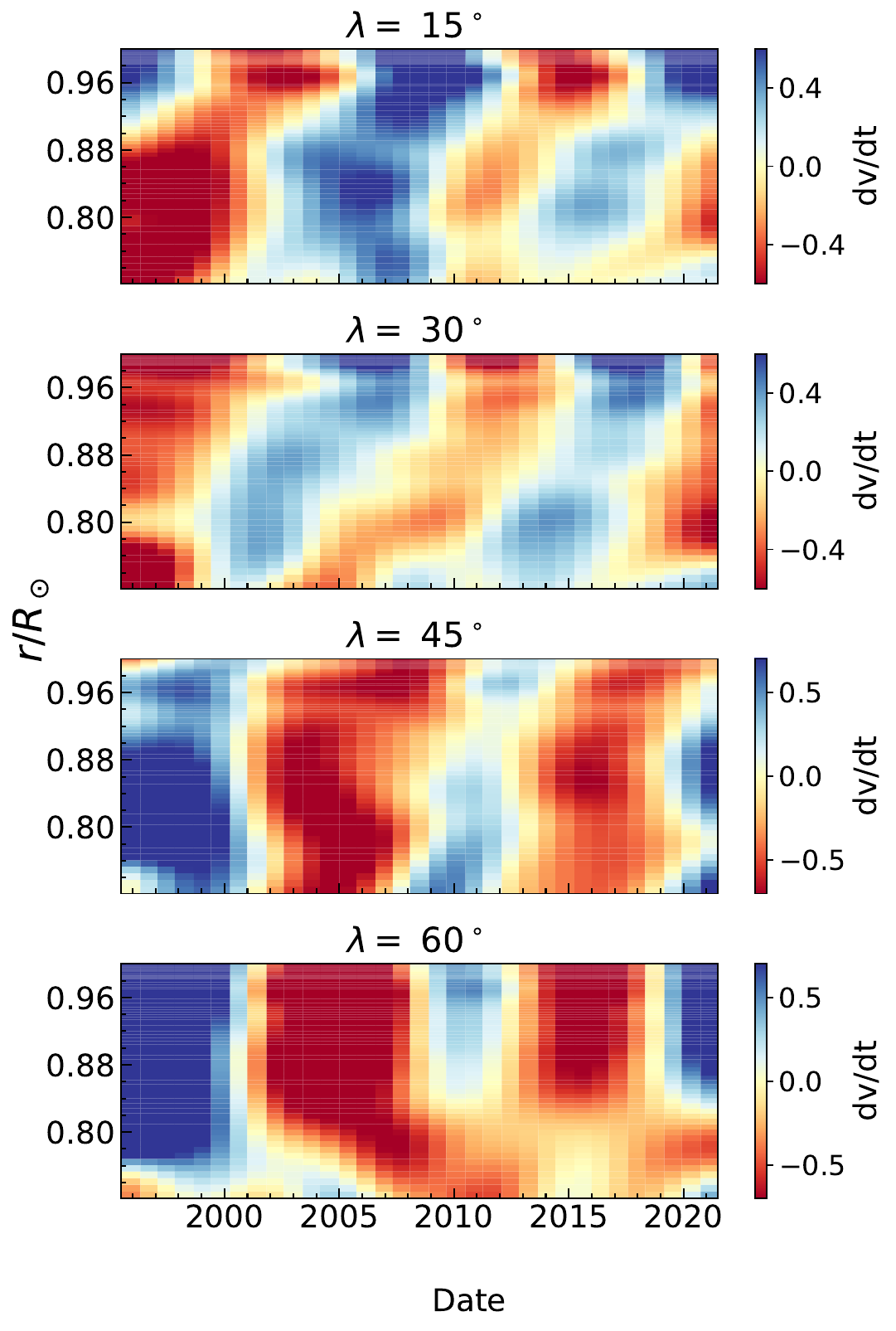}
      \caption{Zonal flow acceleration, derived using the GONG observations, plotted as a function of radius and time for various latitudes 
      ($15^\circ$, $30^\circ$, $45^\circ$, and $60^\circ$). The panels on the left present results derived from $8 \times 72$-day datasets, while the panels on the right correspond to the analysis using $4 \times 72$-day datasets, both from \citet{sylvain23}, derived from GONG observation.}
      \label{fig:dynamo_rad_time_deriv}
  \end{figure} 

\begin{figure}
    \centering
    \includegraphics[scale=0.4]{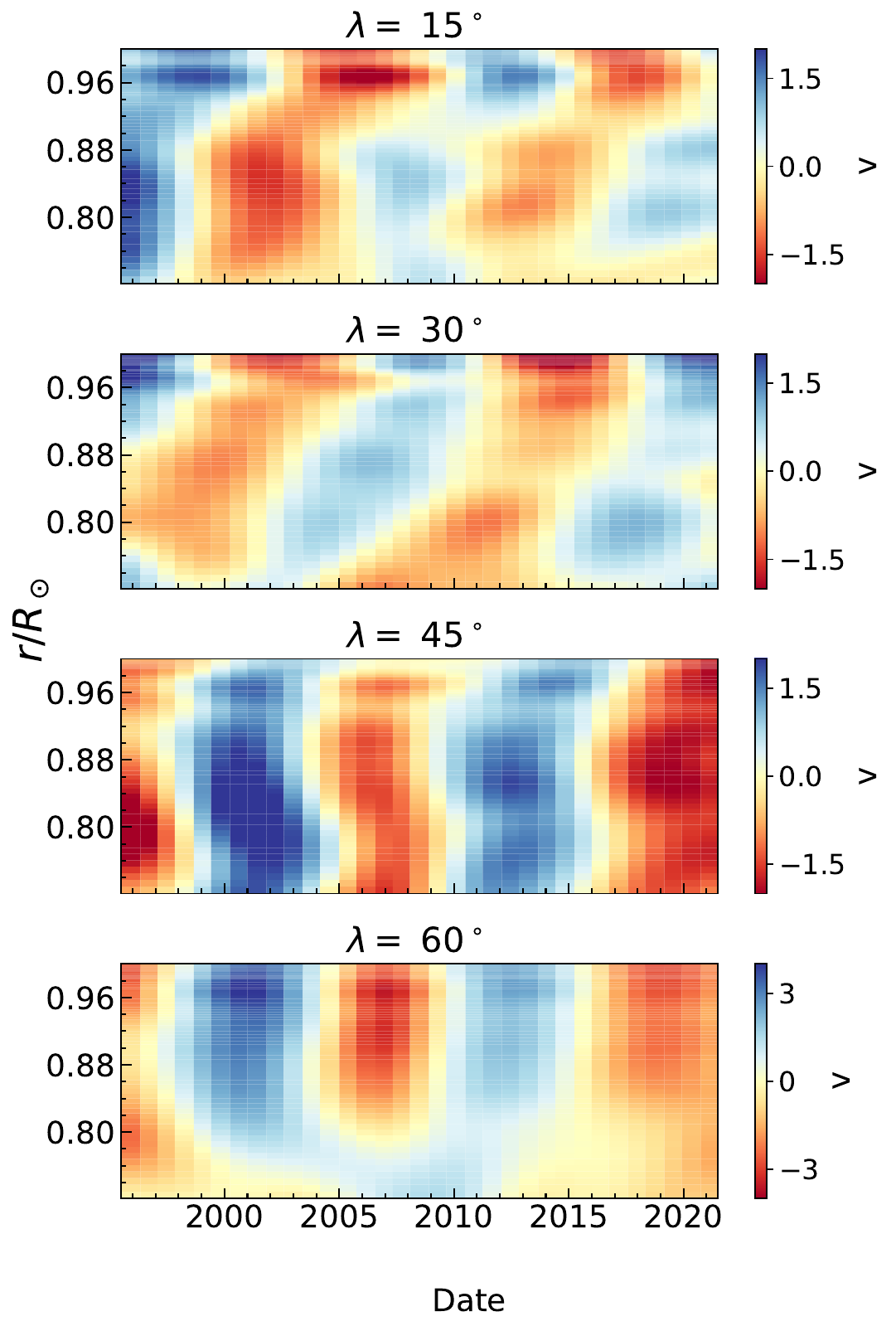}\hspace{1em}\includegraphics[scale=0.4]{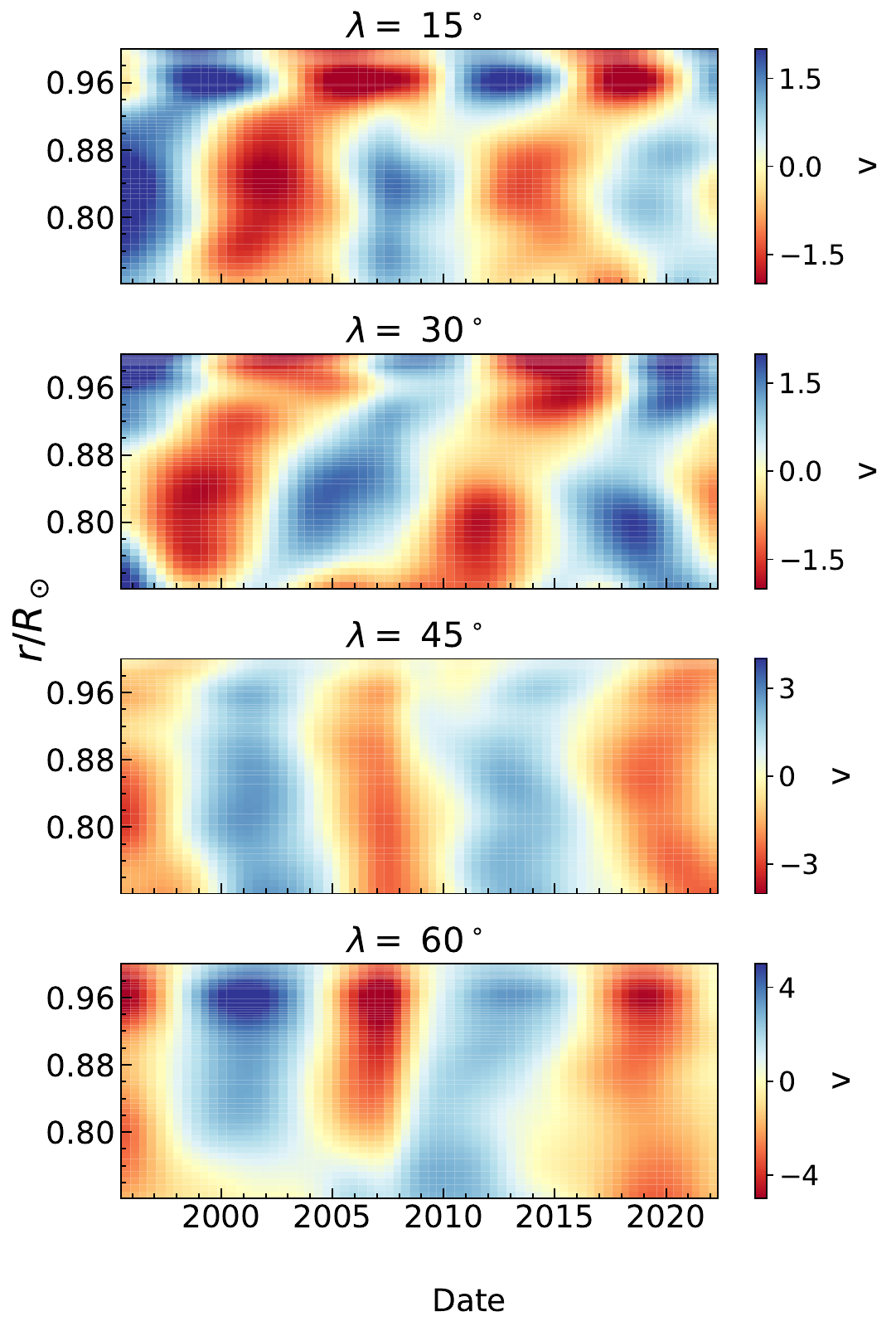}
    \caption{For comparison, the zonal flow is shown using $8 \times 72$-day (left) and $4 \times 72$-day (right) datasets, both from \citet{sylvain23}.}
    \label{fig:dynamo_rad_time_vel}
\end{figure}

We subtract the mean rotation from the measured rotation for each time segment, revealing regions of faster and slower rotation at different latitudes, depending on the phase of the solar cycle. Figure~\ref{fig:dynamo_still} illustrates the evolution of zonal flow over time. We present the zonal flow variations spanning the entirety of Solar Cycle 23. Between 2004 and 2006, a region of slower-than-average rotation emerged near the base of the convection zone at high latitudes, becoming more pronounced until 2010. Between 2010 and 2011, a region of faster-than-average rotation appeared at the same location, intensifying significantly by 2012. Notably, there is a time lag between the emergence of sunspots on the surface and the corresponding changes in zonal flow at the base of the convection zone.   

We plot the time derivative of the zonal flow, i.e., the zonal acceleration, as a function of time and radius in Figure~\ref{fig:dynamo_rad_time_deriv}. A similar plot for the zonal flow itself is shown in Figure~\ref{fig:dynamo_rad_time_vel}. In both figures, we find that from the equator to mid-latitudes, there is a noticeable tilted slope, implying that the pattern, after originating at the base of the convection zone takes about 5–6 years to reach the surface. In contrast, at high latitudes, the pattern is almost vertical, indicating it requires little time to reach the surface.

\begin{figure}
      \centering
      \includegraphics[scale=0.4]{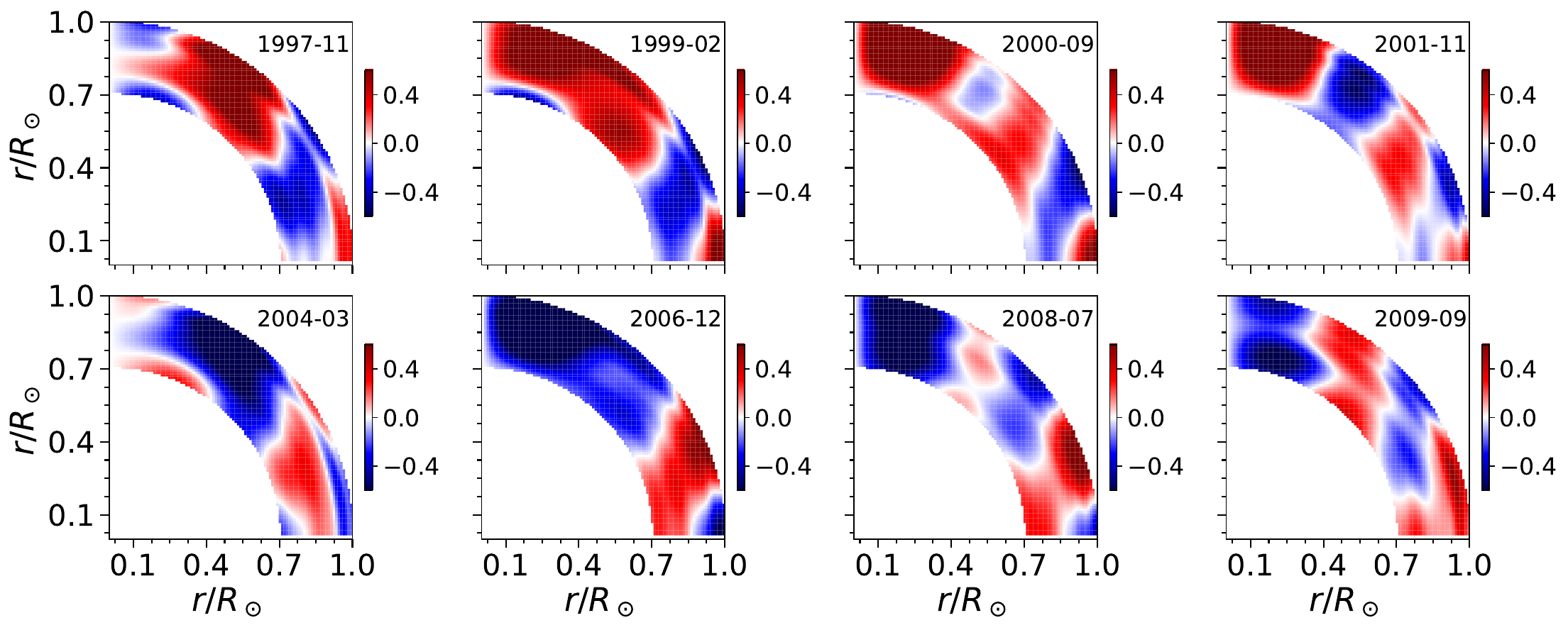}
      \caption{The acceleration of the zonal flow is shown as a function of radius and latitude at various time instances, as indicated in the legend of each panel. Note the difference between Figure \ref{fig:dynamo_still} and this figure—the changes at high latitudes occur at different times, as discussed in the main text. This is due to the differing phase lag between velocity and acceleration. We use $5\times 72$-day datasets from \citet{sylvain23} using GONG observations for the analysis. A video is available online that illustrates the evolution of the zonal flow over the entire time period of our analysis. }
      \label{fig:dynamo_deriv_evol}
  \end{figure}

\begin{figure}     
      \includegraphics[scale=0.4]{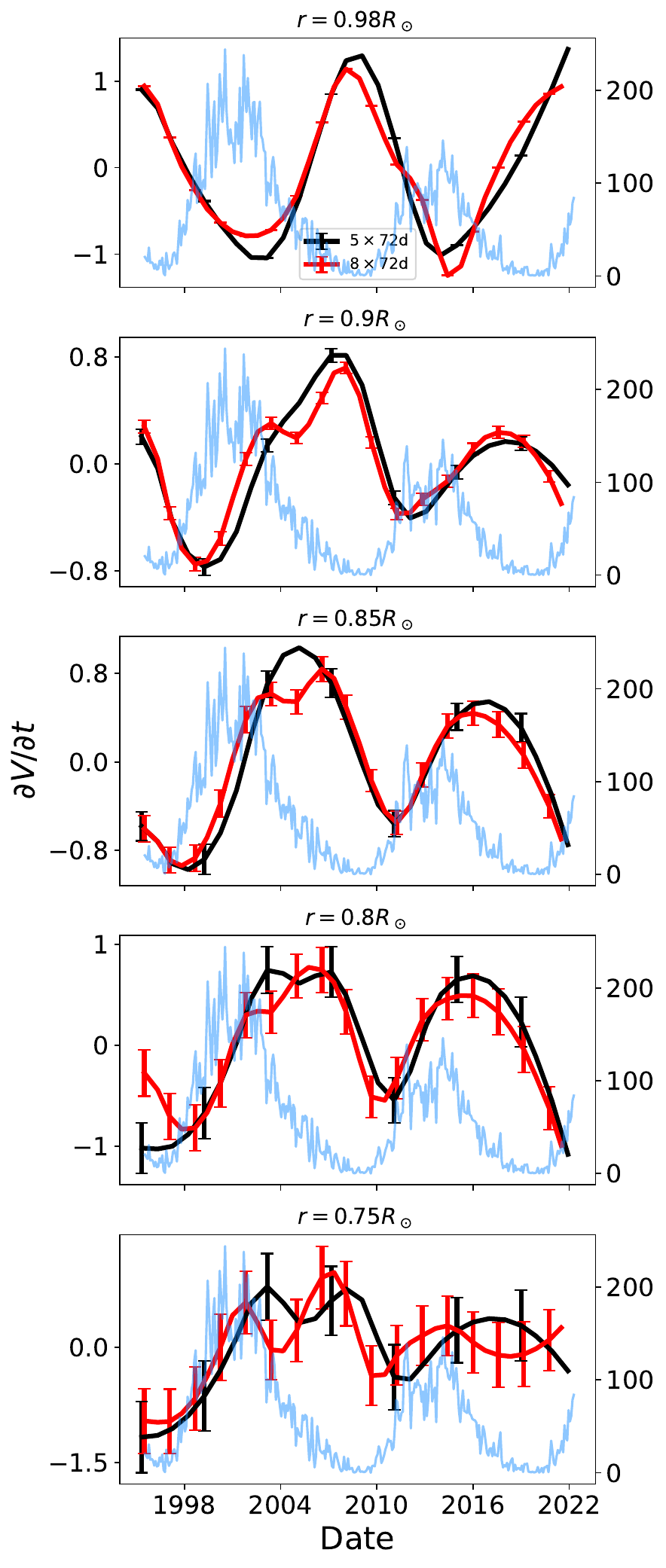}\includegraphics[scale=0.4]{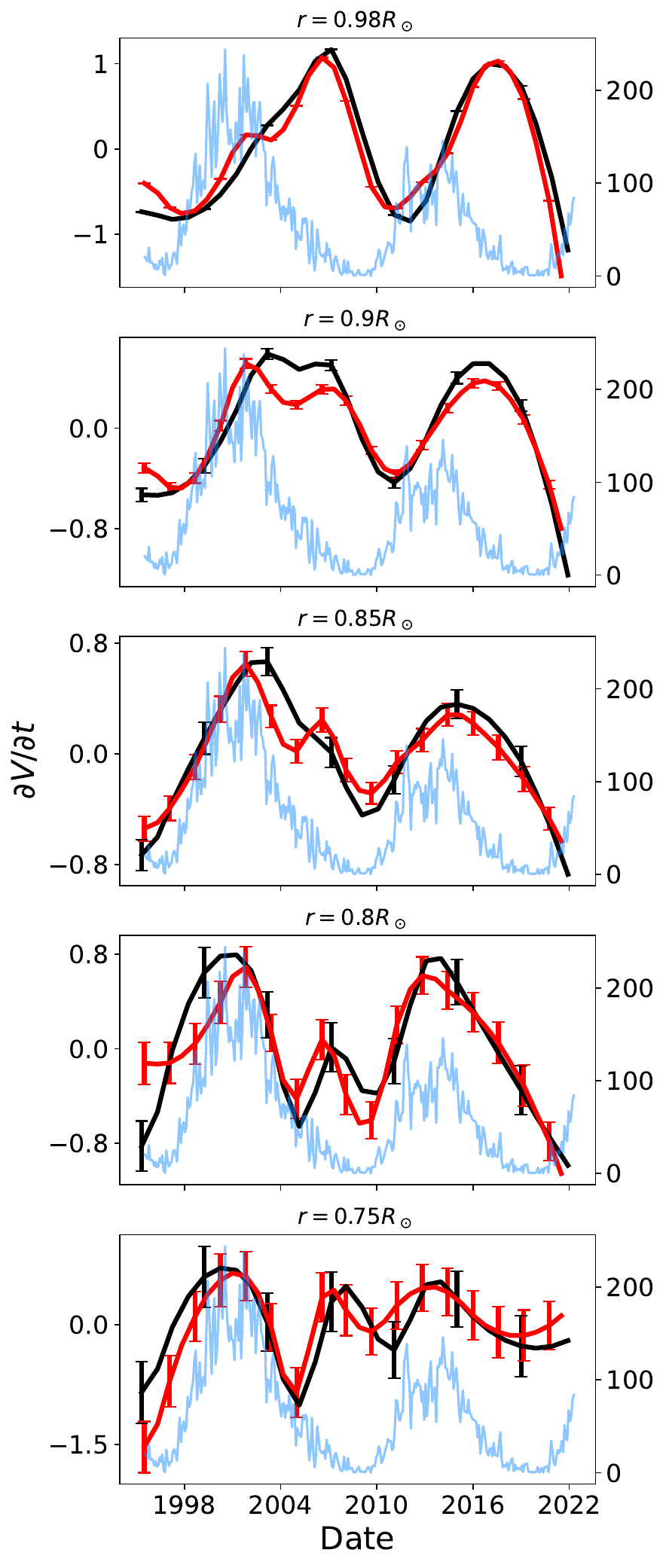}
        \includegraphics[scale=0.4]{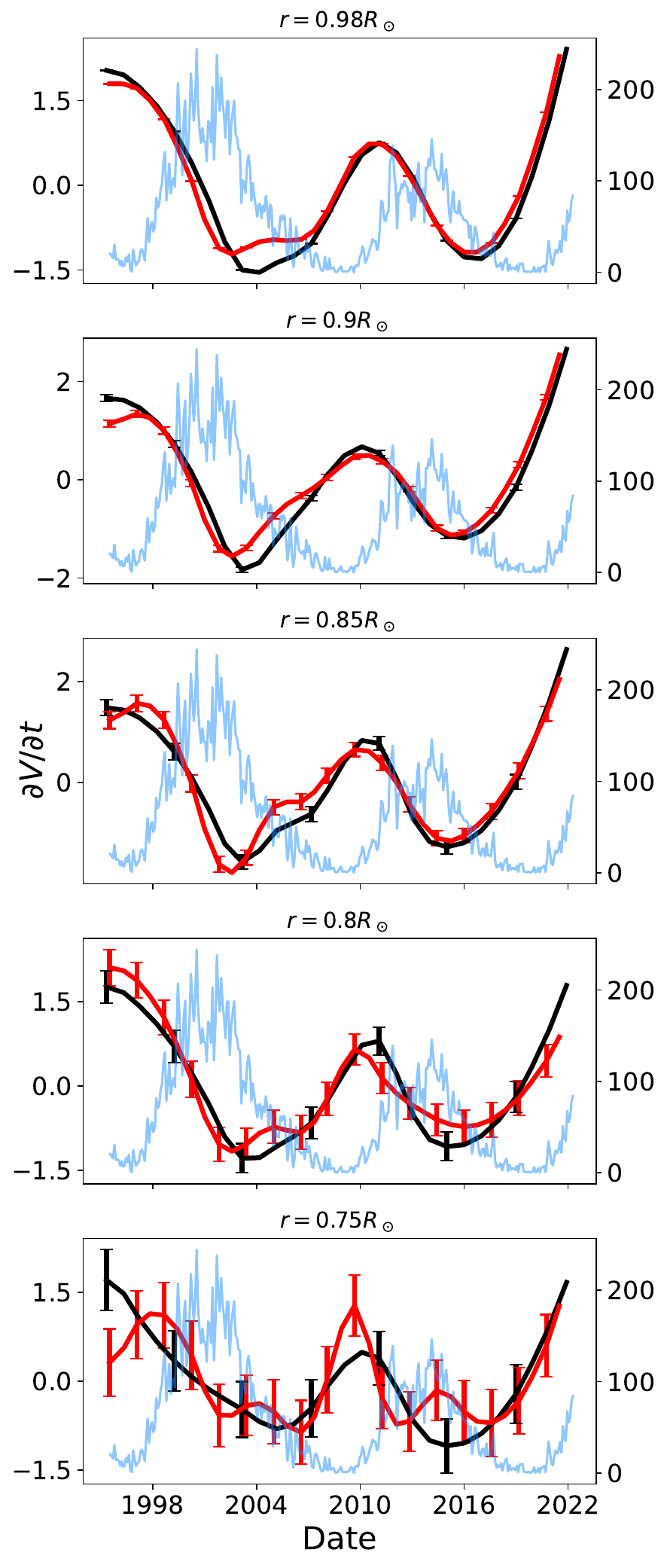}
      \caption{We compare the acceleration of zonal flow, $\dot{V}$, at latitudes $15^\circ$ (left panel), $30^\circ$ (middle panel) and $50^\circ$ (right panel) across multiple depths (as indicated in the title of each panel) using two different GONG dataset lengths: $5 \times 72$-day, and $8 \times 72$-day (black and red curves), both from \citet{sylvain23}. For comparison, we show the corresponding variation of the sunspot number \citep[using solid light blue line][]{schatten78}.}
      \label{fig:dv_tach_comp}
  \end{figure}

\begin{figure}
      %\centering
      %\includegraphics[scale=0.5]{fig/v_tach_compare_50_deg.pdf}
      \includegraphics[scale=0.4]{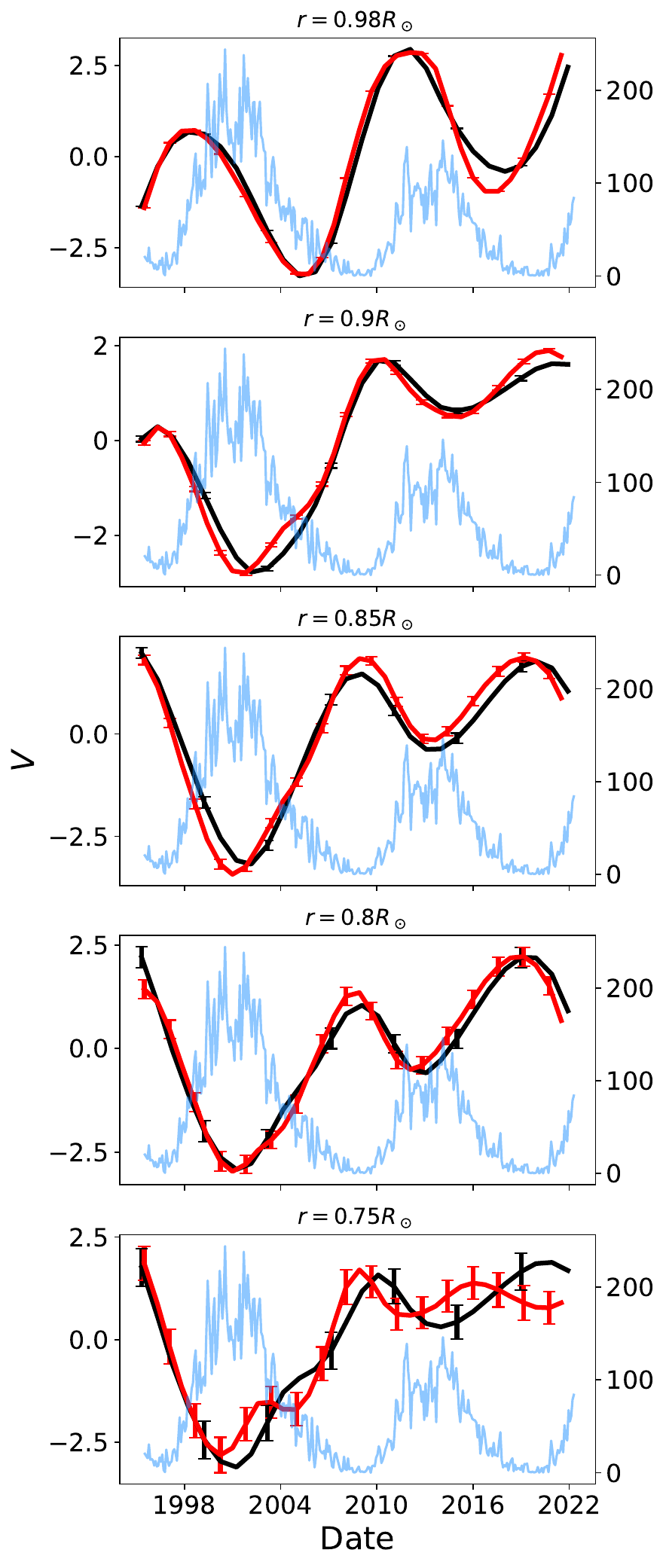}\includegraphics[scale=0.4]{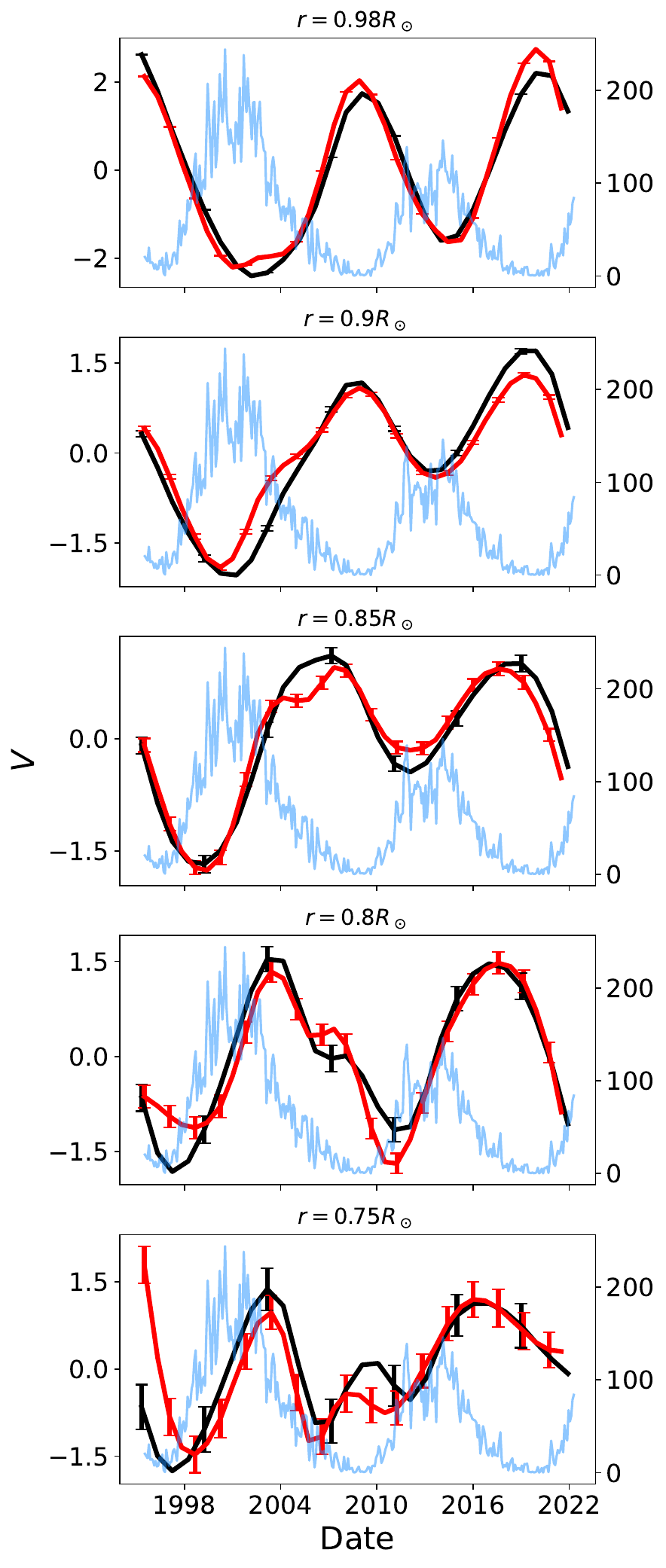}
        \includegraphics[scale=0.4]{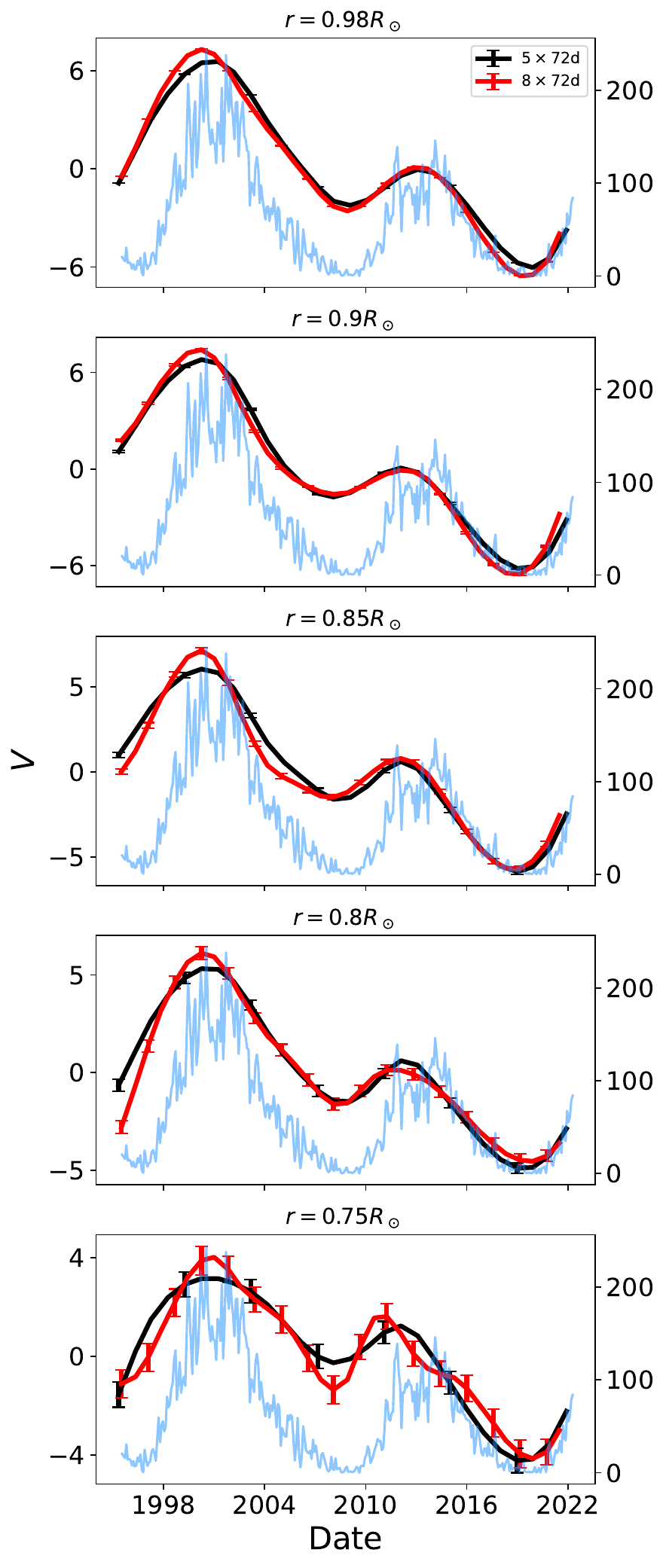}
      \caption{We compare evolution of zonal flow $V(r,\theta,t)$, at latitudes $15^\circ$ (left panel), $30^\circ$ (middle panel) and $50^\circ$ (right panel) across multiple depths (as indicated in the title of each panel) using two different GONG dataset lengths: $5 \times 72$-day, and $8 \times 72$-day (black and red curves), both from \citet{sylvain23}. Sunspot number is shown by solid light blue line.}
      \label{fig:v_tach_comp}
  \end{figure}

Similar to Figure \ref{fig:dynamo_still}, we illustrate the evolution of acceleration in Figure \ref{fig:dynamo_deriv_evol}. Analyzing acceleration is more straightforward because it does not require subtracting the mean, which is necessary when studying the variations in zonal flow. Indeed, the mean can be chosen separately for each cycle -- given that different cycles exhibit varying zonal flow strengths -- hence subtracting a common mean is not ideal for distinguishing clearly different characteristics, such as high-latitude and near-equatorial features. In each case, signatures of dynamo waves are observed in the acceleration of zonal flow. In the first case, the features appear more smoothed, as expected from the analysis of a longer data period. However, since the signature remains visible in both cases, this demonstrates the robustness of the signal. 
We compare acceleration of zonal flow, $\partial V/\partial t = \dot{V}$ and zonal flow itself $V$ at several depths ranging from $0.98 R_\odot$ to $0.75 R_\odot$. This comparison is performed using two datasets of different lengths: $5 \times 72$-days, and $8 \times 72$-days as shown in Figures~\ref{fig:dv_tach_comp} and~\ref{fig:v_tach_comp}. The profile becomes smoother with the longer time series; however, the overall trend of variation is consistently captured in all cases, as shown in the Figures~\ref{fig:dv_tach_comp} and~\ref{fig:v_tach_comp}. We observe phase variations with increasing depth at specific latitudes, such as $15^\circ$ and $30^\circ$. However, at higher latitudes (e.g., $50^\circ$), the phase remains consistent across all depths. This indicates that changes occurring at high latitudes propagate to the surface much faster than those at lower latitudes. The phase lag observed between the zonal flow and its acceleration in earlier figures can also be seen when comparing Figures~\ref{fig:dv_tach_comp} and~\ref{fig:v_tach_comp}.

Focusing solely on the transition from positive to negative values, these two parameters yield different timing for changes near the tachocline. If this pattern serves as an indicator, it could provide insight into key properties of the next solar cycle, such as its onset and strength. However, before drawing conclusions about the solar-cycle behavior based on the torsional oscillations, we must establish a clear correlation between their variations near the tachocline and the magnetic field.  This understanding is crucial for interpreting the solar cycle’s dynamics. A collaborative effort involving modeling is necessary to fully decode these details. An effort in this direction was attempted by \citet{pipin20}. In that study, they highlighted that the sign of the correlation and the time lag depend on the depth and latitude of torsional oscillations, as well as the characteristics of long-term variations in the solar cycle. Of course, more work is needed in that direction to fully understand the relationship between the torsional oscillations and the solar cycle.

 % \begin{figure}
 %      \centering
 %      \includegraphics[scale=0.5]{fig/zf_sylvain_5x72d.pdf}
 %      \caption{Zonal flow is plotted as a function of radius and time for several latitudes, as indicated at the top of each panel.}
 %      \label{fig:dynamo_rad_time}
 %  \end{figure} 

\subsection{Radial Rotation-rate gradient in NSSL}
\begin{figure}
     \includegraphics[scale=0.45]{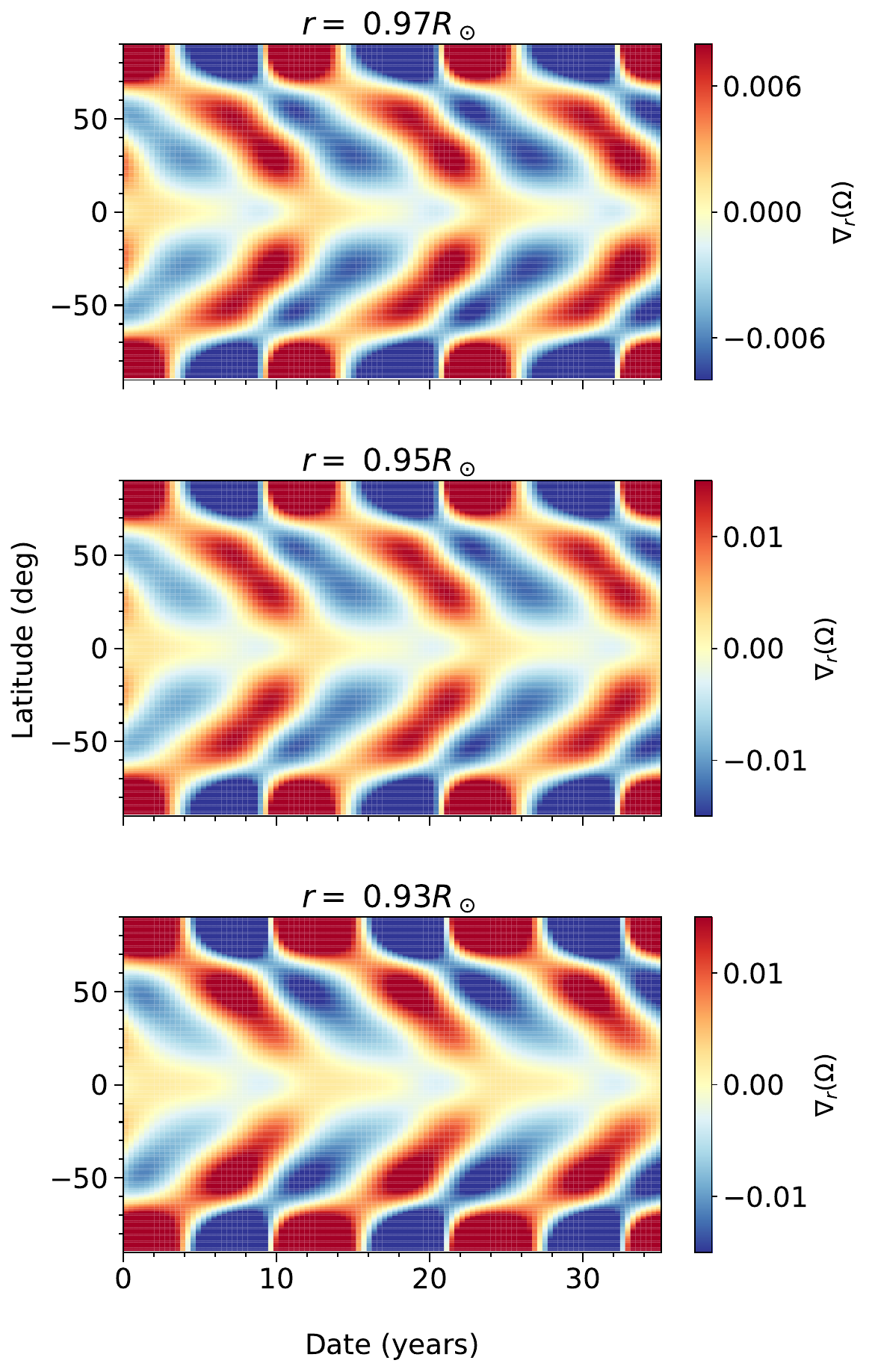}\includegraphics[scale=0.45]{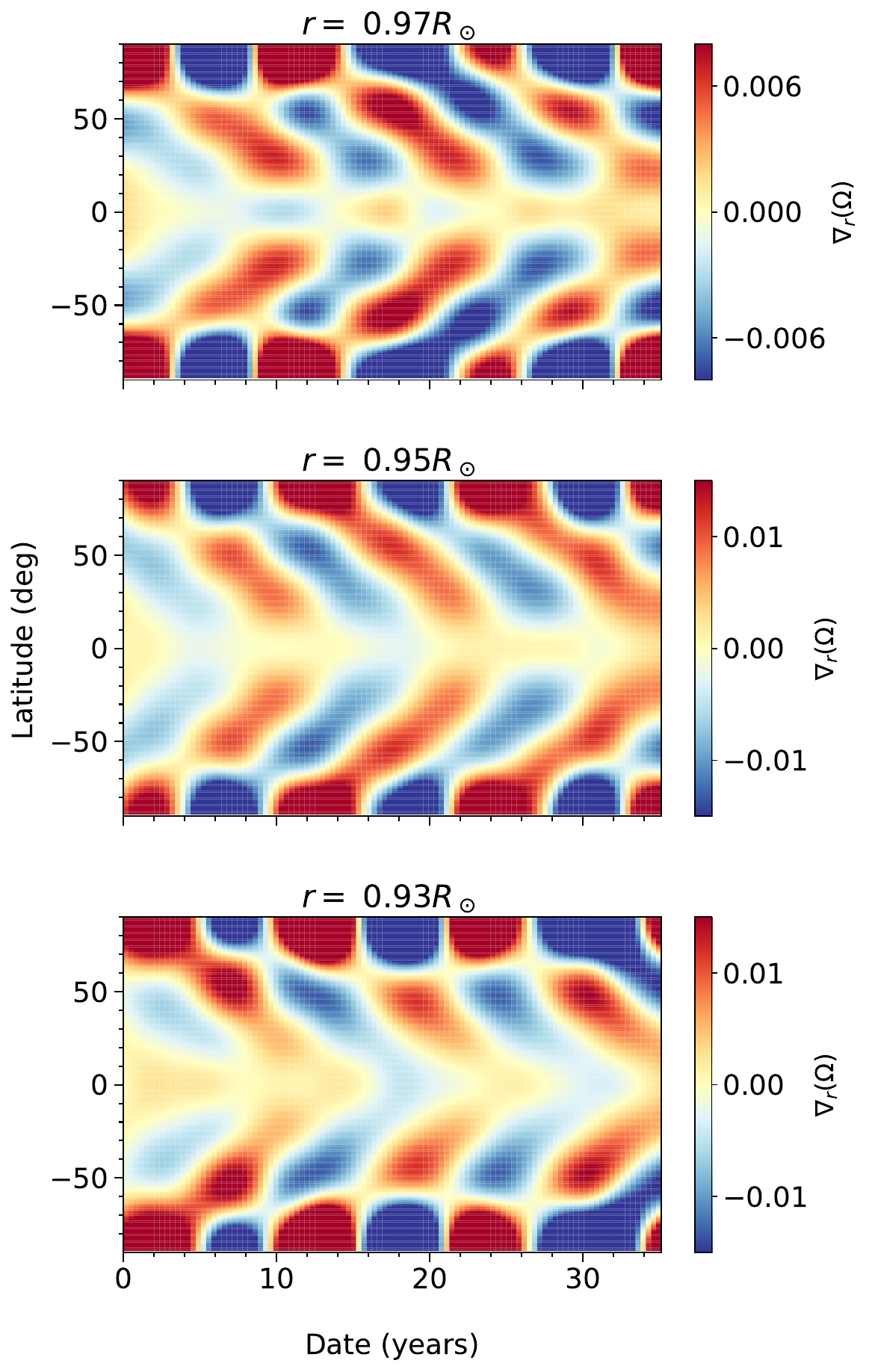}
     \caption{The left panels show the variation in the dimensionless radial gradient, $\nabla_r(\Omega)$, as obtained from the dynamo model of \citet{pipin20}. The right panels display the corresponding results derived from the inversion, using the dynamo model profile as input. Results are presented for several depths, as indicated in the title of each panel.}
     \label{fig:rad_grad_syn}
 \end{figure}

 Using the dynamo model from \citet{pipin20}, we calculate the dimensionless radial gradient, $\nabla_{r}(\Omega)$. We subtract its mean temporal value and plot the result in Figure \ref{fig:rad_grad_syn}. We find that the equatorward branch originates at very high latitudes and then migrates toward the equator. We then use the rotation profile from the same dynamo model to perform forward modeling, generate the corresponding a-coefficients, carry out an inversion, and estimate the dimensionless radial gradient, again subtracting the mean temporal profile. The results are shown in the right panel of Figure \ref{fig:rad_grad_syn}, where we find a reasonable match between the original profile and the inverted profile, validating our methodology. Next, we analyze the radial gradient of rotation based on the inversion of the observed a-coefficients.
 
\begin{figure}
      \centering
      \includegraphics[scale=0.4]{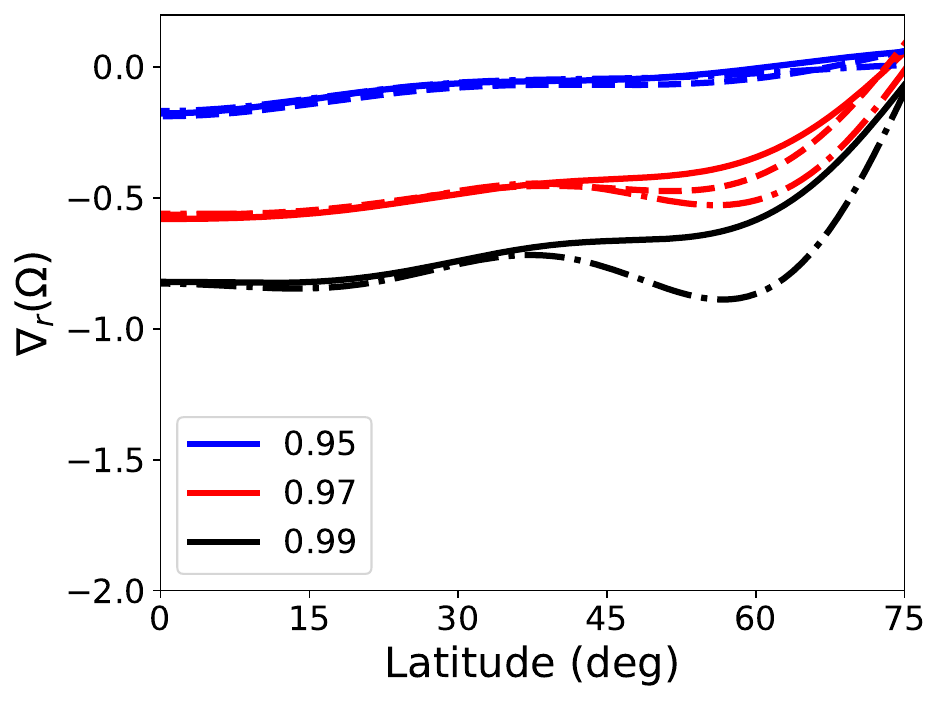} \hspace{1em}\includegraphics[scale=0.4]{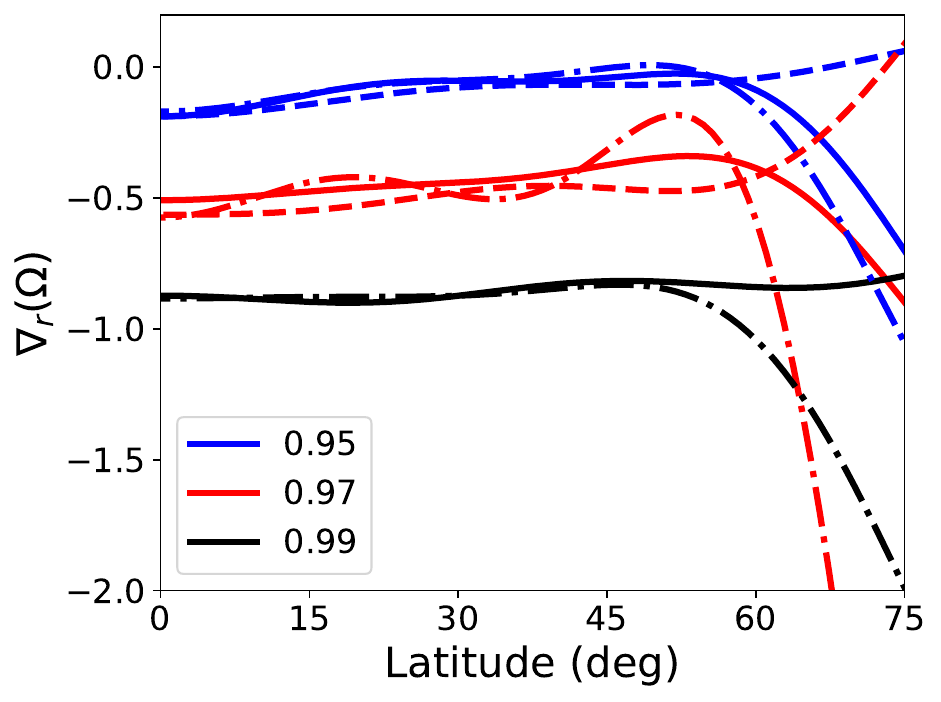}\\
      \includegraphics[scale=0.4]{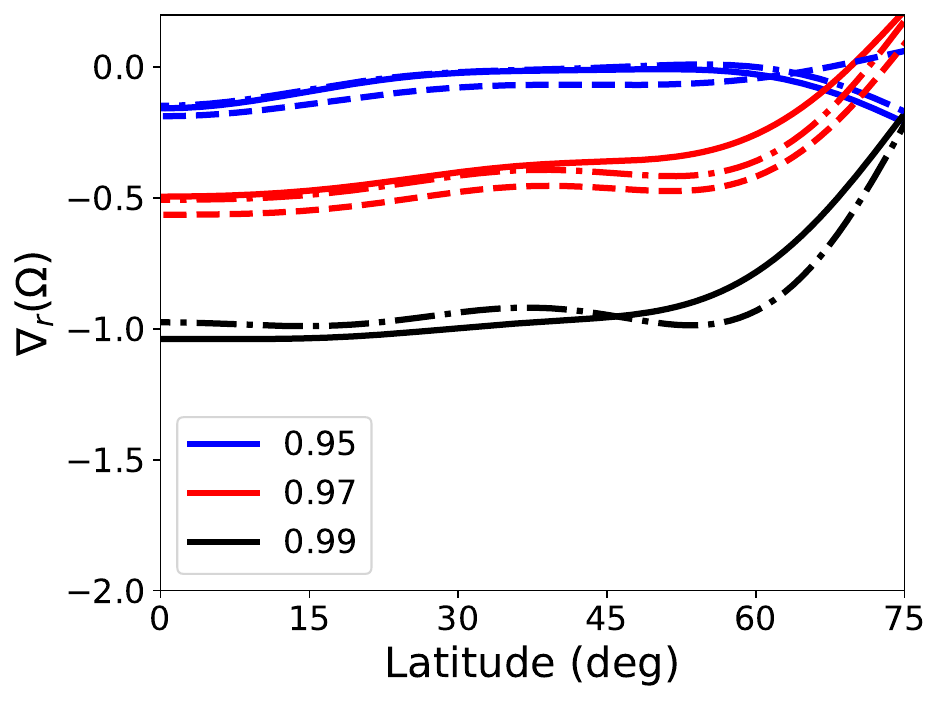}\hspace{1em}\includegraphics[scale=0.4]{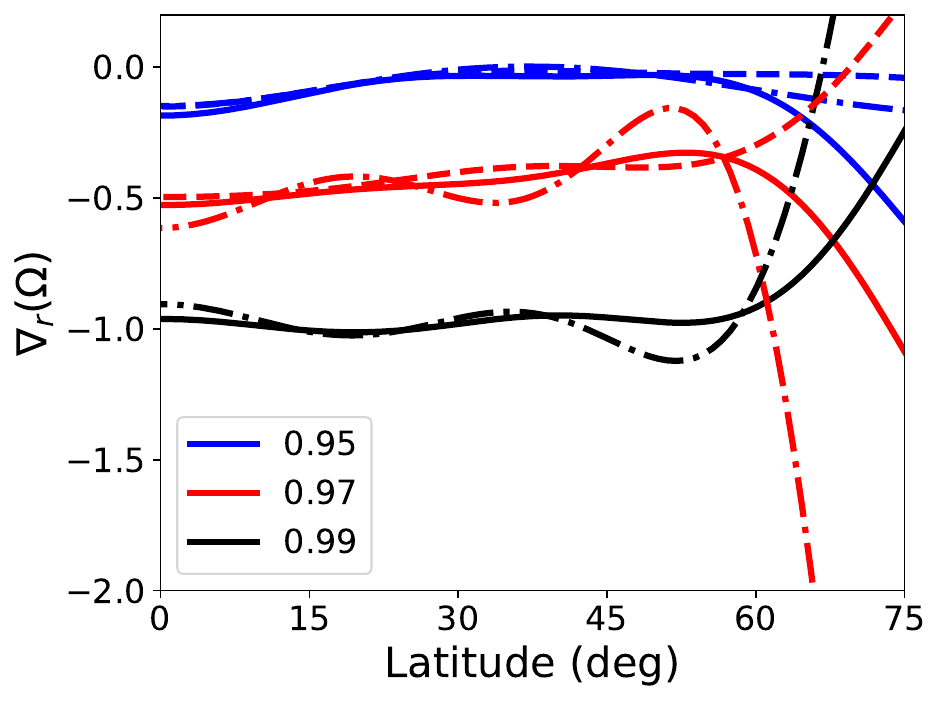}
      \caption{The mean radial gradient of rotation, $\nabla_r(\Omega)$ is shown as a function of latitude at three different depths—0.95, 0.97, and 0.99 $R_\odot$ —represented by blue, red, and black curves, respectively. The data are taken from GONG, MDI, and HMI observations, indicated by dashed, dash-dotted, and solid lines, respectively. Results from GONG at the depth $0.99 R_\odot$ are not shown, as the maximum harmonic degree for GONG datasets is limited to $150$. The left panels show results from the analysis of the datasets used in \citet{sylvain23}, with the top panel corresponding to the analysis of $1\times 72-$day datasets and the bottom panel to $5\times 72-$day datasets. The right panels show corresponding results based on datasets processed using the JSOC pipeline, except for GONG, for which we use the datasets from \citet{sylvain23}. The error bars on all plots are very small when using a time series composed of $5\times 72$-day data sets.}
      \label{fig:dln_dr_lat}
  \end{figure}

 We use GONG datasets with a maximum harmonic degree of $\ell = 150$, which may lead to some discrepancies near the surface. However, we still present results near the surface from three different instruments MDI, HMI, and GONG in Figure \ref{fig:dln_dr_lat} for three different depths, all within the NSSL. As shown in Figure \ref{fig:dln_dr_lat}, the absolute value of $\nabla_r(\Omega)$ decreases with radius. Near the surface $\approx 0.99 R_\odot$, GONG results deviate slightly due to the dataset's limitation in providing frequency splitting for harmonic degrees $\ell > 150$, which is insufficient to resolve the near-surface layers. The mean value of $\nabla_r(\Omega)$ at high latitudes ($>60^\circ$) shows inconsistencies between instruments, as evident in Figure \ref{fig:dln_dr_lat}. The differences at high latitudes exceed significantly, indicating potential systematics in the frequency measurements provided by \citet{sylvain23} and the JSOC pipeline. The radial gradient of angular velocity, $\nabla_{r}(\Omega)$, remains nearly constant from the equator to mid-latitudes, consistent with the findings of \citet{barekat2014, antia2022}, but in contrast to \citet{corbard2002}, who reported a decrease with latitude.  
 
\begin{figure}
 \centering
     \includegraphics[width=0.4\linewidth]{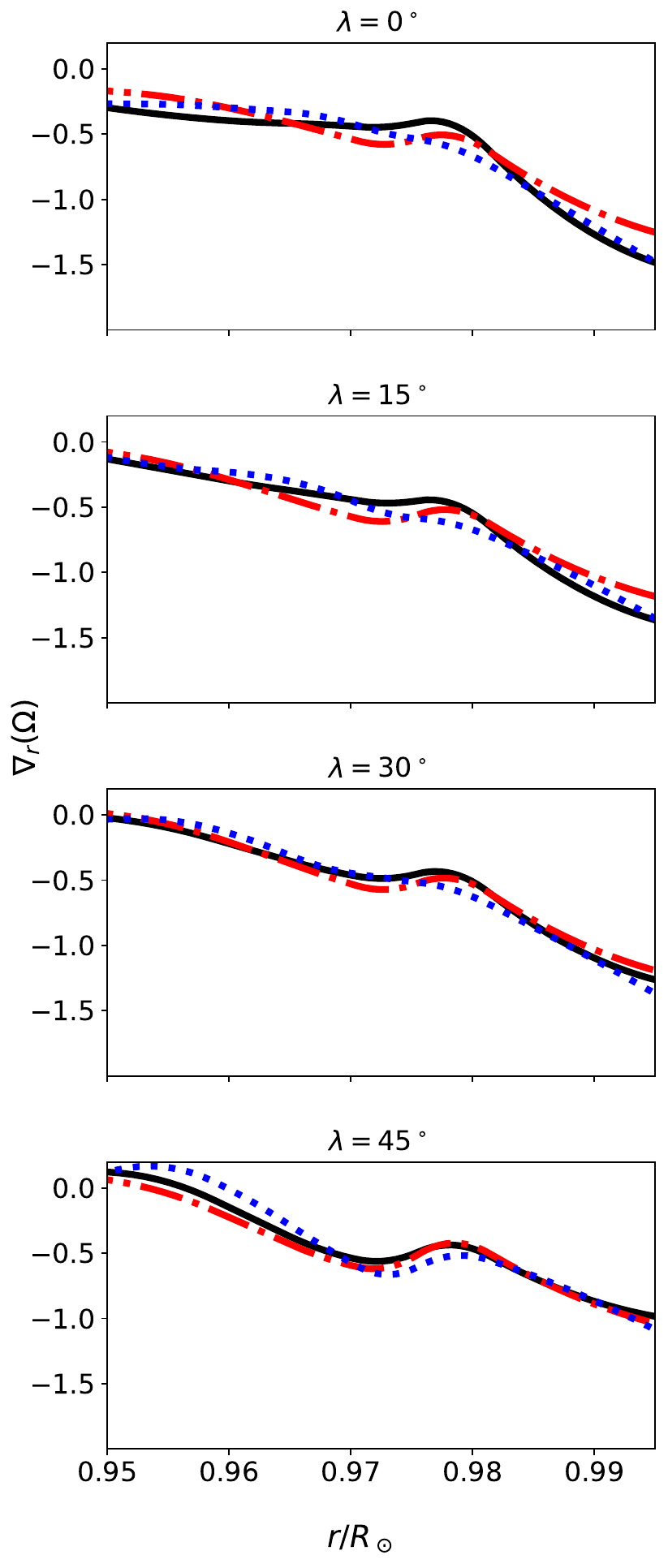}\hspace{1em}\includegraphics[width=0.4\linewidth]{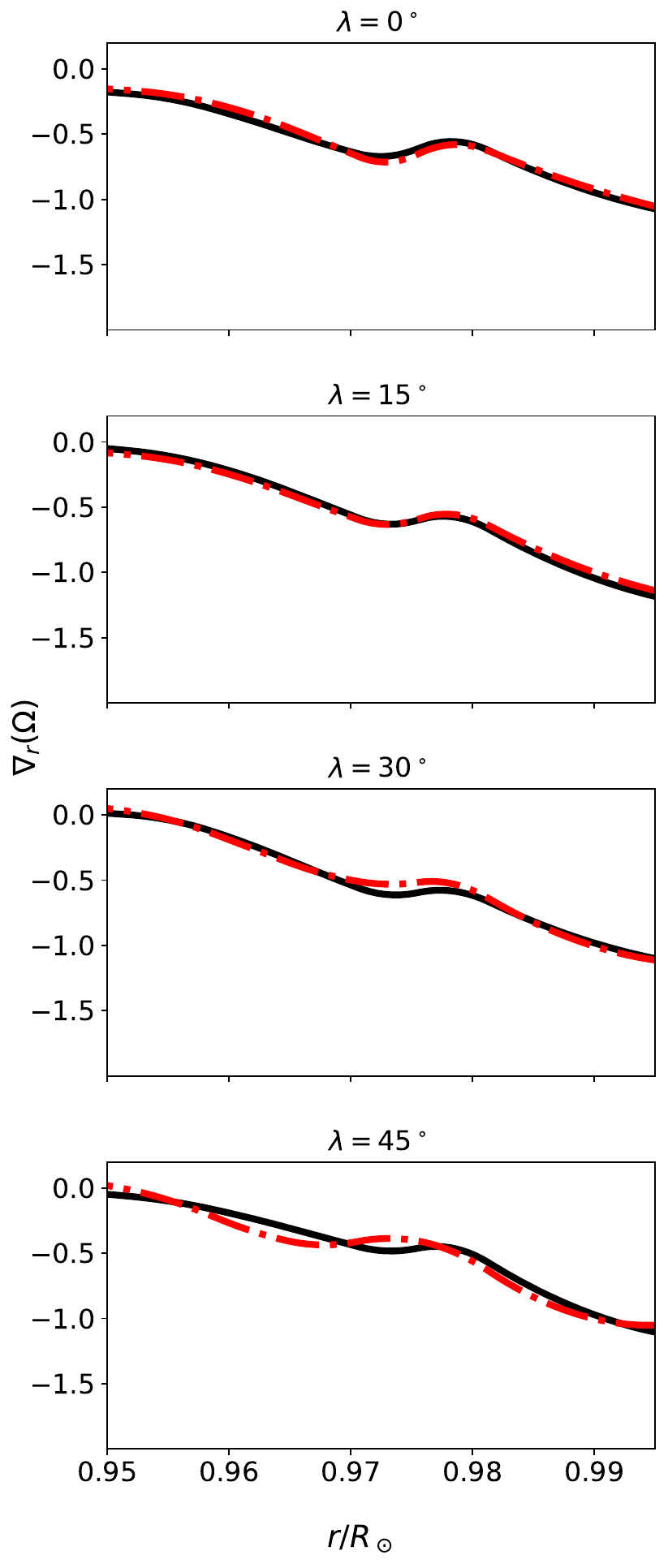}
     \caption{We plot $\nabla_r(\Omega)$ as a function of radius at several different latitudes (indicated in the title of each panel). The left panel shows the results using the data series from \citet{sylvain23}, while the right panel shows the results using the data series from the JSOC pipeline. In the right panel, we compare only HMI (black solid line) and MDI (red dash-dot line) results, whereas in the left panel, we also include results from GONG observations (blue dotted line). We use $5\times 72$-day time series for all the plots.}
     \label{fig:dln_radius}
 \end{figure}

 We show how the temporal mean of $\nabla_r(\Omega)$ varies with depth for different latitudes in Figure \ref{fig:dln_radius} for all the instruments. We find that there is a bump at depth $\sim 14$ Mm below the surface or around depth $~0.98 R_\odot$. A similar bump was also observed by \citet{komm2023,cristina2024}. 
 Near this depth, the HeII ionization zone transitions into the HeI and HI ionization zones. This change in ionization leads to a sharp decrease in the adiabatic index, $\Gamma_1$, toward the surface. We  speculate that the presence of these transition layers is responsible for the bump observed at this depth. Figure \ref{fig:dln_radius} shows that the value of the dimensionless radial gradient increases from the surface to the deeper layers. Its value is approximately $-1$ near the surface and increases with depth. \citet{irina2023} performed a radiative magneto-hydrodynamic simulation of a local box at latitude $30^\circ$, with a box size of 80 Mm in width and 25 Mm in depth, and found a qualitative match between the simulated and observed radial gradients of rotation. \citet{cristina2024} found that the radial gradient shows significant variation down to 5 Mm below the surface -- a region very close to the surface that cannot be investigated using our analysis. In the future, if we can analyze very high harmonic degrees ($>300$), it may become possible to investigate these very shallow layers of the Sun.
  
\begin{figure}
     \centering
     \includegraphics[width=0.65\linewidth]{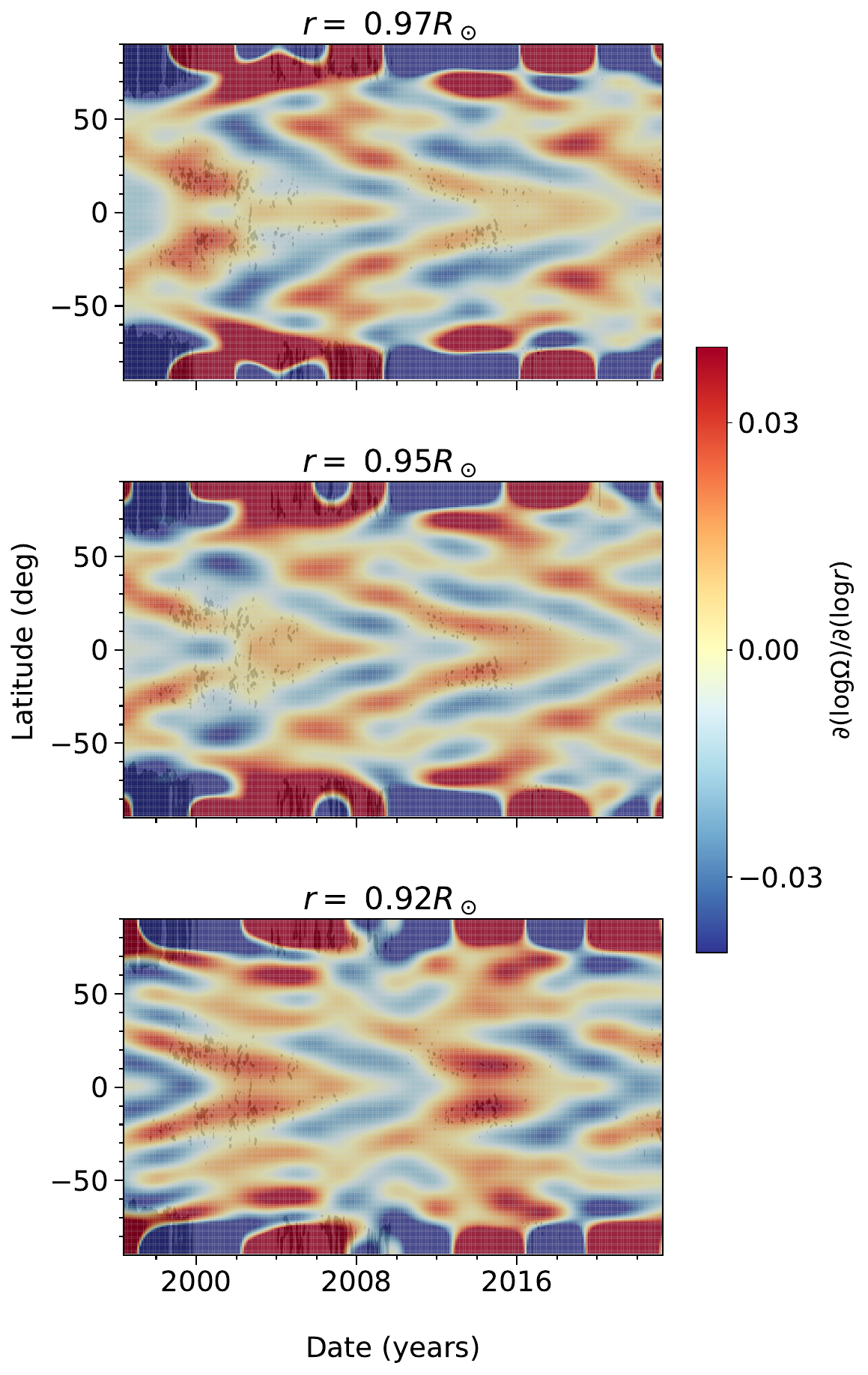}
     \caption{The variation of angular velocity gradient,  $\nabla_r(\Omega)$ as a function of time and latitude for three different depths as indicated at the top of each panel. Sunspot locations are shown as gray dots overlaid on the plot to facilitate comparison with the magnetic butterfly diagram. For this analysis, we use the $5\times 72$-day datasets from \citet{sylvain23} derived from GONG observations.} 
     \label{fig:dln_domega_dr}
 \end{figure}

 Next, we subtract the temporal mean of $\nabla_r(\Omega)$ from the results obtained for each 72-day time series analysis. This reveals a torsional oscillation-like signature. The migration of higher-than-average values from mid-latitudes toward the equator is clearly visible in Figure \ref{fig:dln_domega_dr}, whereas the mid-latitude to polar migration is less distinct, unlike the patterns observed in torsional oscillation. The mid-latitude to polar branch appears more complex, while the mid-latitude to equator branch aligns well with the magnetic field migration seen in the sunspot butterfly diagram. Notably, there is a striking similarity between the model predictions shown in Figure \ref{fig:rad_grad_syn} and the observed results in Figure \ref{fig:dln_domega_dr} regarding the variation of $\nabla_r(\Omega)$.   
 
  
 % \begin{figure}
 %     \centering
 %     \includegraphics[scale=0.4]{fig/dln_dr_lat_radius_MDI.pdf}\includegraphics[scale=0.4]{fig/dln_dr_lat_radius_HMI.pdf}\includegraphics[scale=0.4]{fig/dln_dr_lat_radius_GONG.pdf}
 %     \caption{The angular velocity gradient $\nabla_r(\Omega)$ as a function of radius and latitude, derived from the analysis of MDI (left), HMI (middle), and GONG (right) datasets. We use data from \citet{sylvain23} for this plot}
 %     \label{fig:dln_rad_lat}
 % \end{figure}

 \section{Discussion and Conclusion}
 We have developed a time-dependent inversion method using cubic B-splines in both radius and time for the first time. Previous inversions were performed independently for each time segment, with additional temporal smoothing applied to reduce variations over time. With this new approach, separate smoothing is no longer required; instead, smoothing is controlled directly during inversion by adjusting the parameter $\lambda_t$ in Equation \ref{eq:chi}. Near the tachocline, the optimal averaging period for a smooth solution is uncertain. This concern is eliminated in the time-dependent inversions, as the smoothing parameters inherently manage the smoothing process in radius and time. 

We also present the dimensionless rotation gradient $\nabla_{r}(\Omega)$, which is approximately $-1$ near the surface and gradually approaches zero toward the middle of the convection zone, consistent with the findings of \citet{antia2022,komm2023}. Deviations from the mean profile exhibit a torsional oscillation-like behavior similar to patterns observed in traditional torsional oscillation measurements, though the high-latitude branch appears more complex. We find that the mean $\nabla_{r}(\Omega)$ exhibits some discrepancies between different frequency measurement methods at high latitudes ($>60^\circ$), while showing good agreement at lower latitudes ($<60^\circ$). We observe a bump in the radial rotation gradient near $0.95 R_\odot$, consistent with earlier findings by \citet{komm2023} based on ring-diagram analysis. Since global helioseismology cannot accurately probe regions very close to the solar surface, we refer interested readers to the work of \citet{cristina2024}, whose ring-diagram analysis extends to layers just beneath the surface. Their results show that the radial rotation gradient changes significantly above $0.995 R_\odot$. The origin of the NSSL remains an active area of research. For instance, global magnetohydrodynamic (MHD) simulations aiming to self-consistently reproduce the NSSL as observed by helioseismology. \citet{matilsky_2019} were able to reproduce the NSSL, but only at low latitudes ($<15^\circ$). \citet{kitchatinov_2016} suggested that the shear in this layer can be sensitive to the subsurface magnetic field and could serve as a probe for sufficiently strong fields, on the order of 1 kG. Our results on the subsurface radial gradient, therefore, can provide useful constraints on the subsurface magnetic field.     

We illustrate the evolution of dynamo waves in Figure \ref{fig:dynamo_still}. A distinct pattern emerges at high latitudes $50^\circ$---\, $60^\circ$, with one branch migrating toward the equator and the other toward the polar region. A similar pattern can be observed in the plot of zonal flow acceleration in Figure \ref{fig:dynamo_deriv_evol}. It highlights the importance of analyzing different properties of the zonal flow to gain a deeper understanding of the solar dynamo and its interaction with zonal flows. We performed inversions of frequency-splitting data from different time intervals and found that signatures of dynamo waves were visible in all cases, highlighting the robustness of the signal. Our analysis includes data from all three instruments and various techniques used to measure frequency splitting. In our previous work \citep[][]{mandal24_dw}, we examined data from the JSOC and GONG pipelines. In this study, we analyze data from \citet{sylvain23} over different time periods, along with datasets from the JSOC pipeline. Both zonal flow and its acceleration exhibit a dynamo wave-like signature. In all cases, the dynamo wave signature remains evident, unambiguously confirming the original detection by \citet{sasha2019}. We find that zonal acceleration and torsional oscillations exhibit a phase difference. They undergo a sign change at high latitudes ($~60^\circ$), with one branch migrating toward the equator while the other moves to higher latitudes. If the sign change in these parameters serves as an indicator of the next solar cycle, their phase lag would result in different timing predictions. Therefore, understanding the correlation between these parameters and the solar cycle is essential for drawing meaningful conclusions about the nature of the next cycle. \citet{pipin20} conducted modeling analyses to explore this connection, suggesting that zonal flow acceleration at high latitudes near the tachocline correlates with the strength of the upcoming solar cycle. Our future work will focus on establishing a correlation between these two parameters and the observed magnetic fields at the surface. Since modeling provides greater control over various parameters—unlike observations, which cannot directly probe magnetic fields below the subsurface layers—we require specific testable quantities. Guidance from modeling studies will be essential in identifying the key link between torsional oscillation and its link to solar activity. 

\section*{}
\begin{flushleft}
\textit{Acknowledgments}: This work was partially supported by NASA grants 80NSSC20K1320, 80NSSC20K0602, and 80NSSC22M0162.
\end{flushleft}

 \pagebreak
 \setlength{\parskip}{0pt}
 
 \appendix
 \section{Methods for Time Dependent Inversion}\label{sec:app_inv}
 In global helioseismology, we generally perform inversion for each 72 days separately to study the evolution of differential rotation with the solar cycle. We smooth the inversion results over time to get smooth variations of the solar differential rotation. Here, we perform inversion in both radius and time so that smoothing in time is included in the inversion procedure. This will capture all the time-varying information. The advantage of this approach is that the smoothing in time can be different at different depths in the convection zone. To do that, we solve the misfit function as shown in Equatuon~\ref{eq:chi} with second derivative smoothing in radius and first derivative smoothing in time.
 We use cubic B-splines for our work for both time and radius and control the regularization by two parameters in Equation \ref{eq:chi} $\lambda_r$ and $\lambda_t$. We expand $w_s(r,t)$ in B-spline basis functions both in radius and time. We choose the second-derivative smoothing in radius and the first-derivative smoothing in time. 

\begin{equation}
    w_{s}(r;t_{k})=\sum_{i}c_{j,i}B_{j}(t_{k})B_{i}(r),
    \label{eq:wst_app}
\end{equation}
Equation \ref{eq:wst_app} can be used to compute the velocity profile as follows
\begin{align*}
V(r,\theta,t) & =\sum_{s}w_{s}(r,t)\hat{r}\times\nabla_{h}Y_{s0}(\theta)\\
\frac{dV(r,\theta,t)}{dt} & =\sum_{s}\sum_{i,j}c_{j,i}\frac{dB_{j}(t)}{dt}B_{i}(r)\hat{r}\times\nabla_{h}Y_{s0}(\theta)\\
a_{s}(n,\ell;t_{k}) & =\int\mathcal{K}_{n,\ell}w_{s}(r;t_{k})r^{2}dr\\
 & =\int\mathcal{K}_{n,\ell}\sum_{i,j}c_{j,i}B_{j}(t_{k})B_{i}(r)r^{2}dr\\
 & =\sum_{i,j}c_{j,i}\int\mathcal{K}_{n,\ell}B_{j}(t_{k})B_{i}(r)r^{2}dr\\
\end{align*}

\begin{figure}
    \centering
    \includegraphics[scale=0.35]{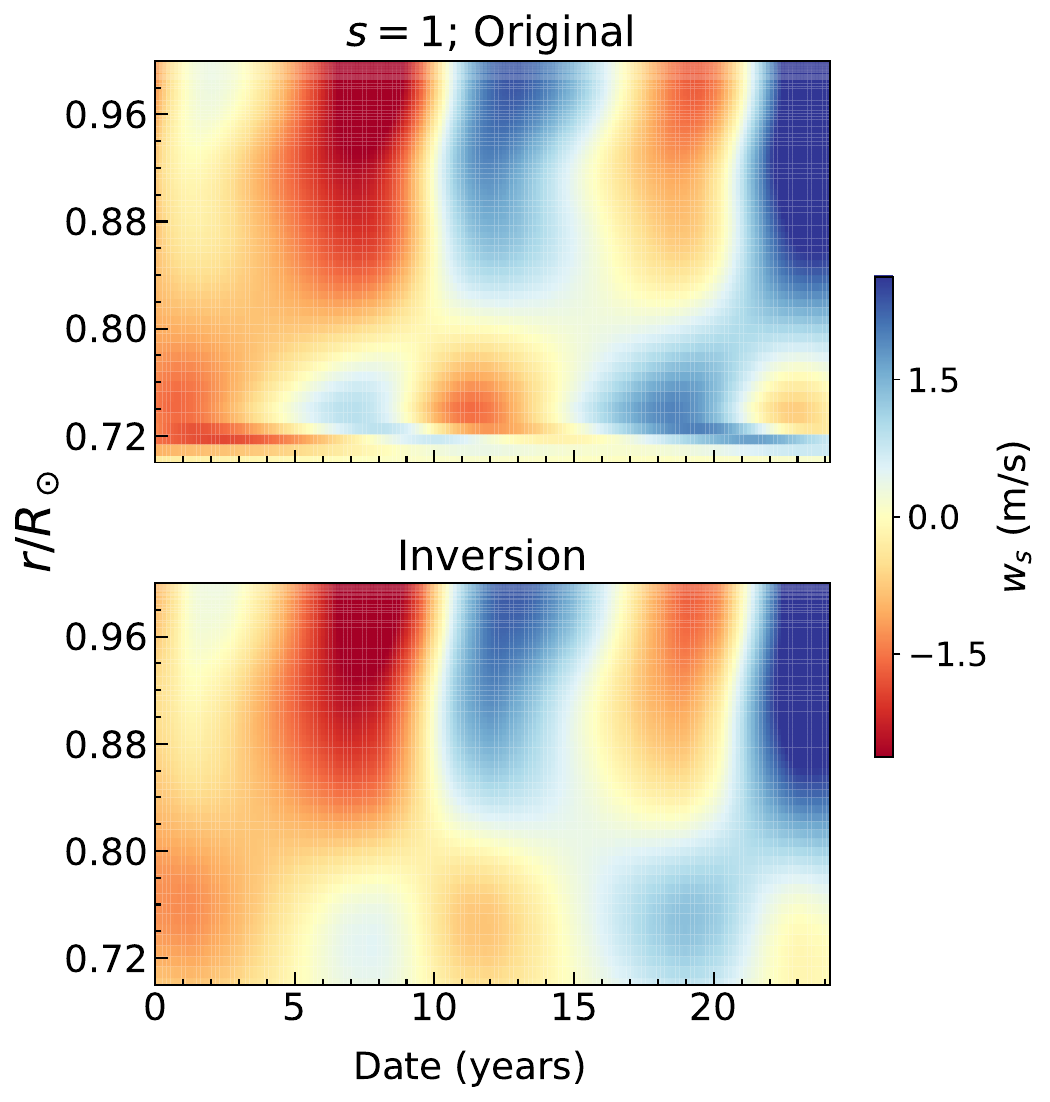}\includegraphics[scale=0.35]{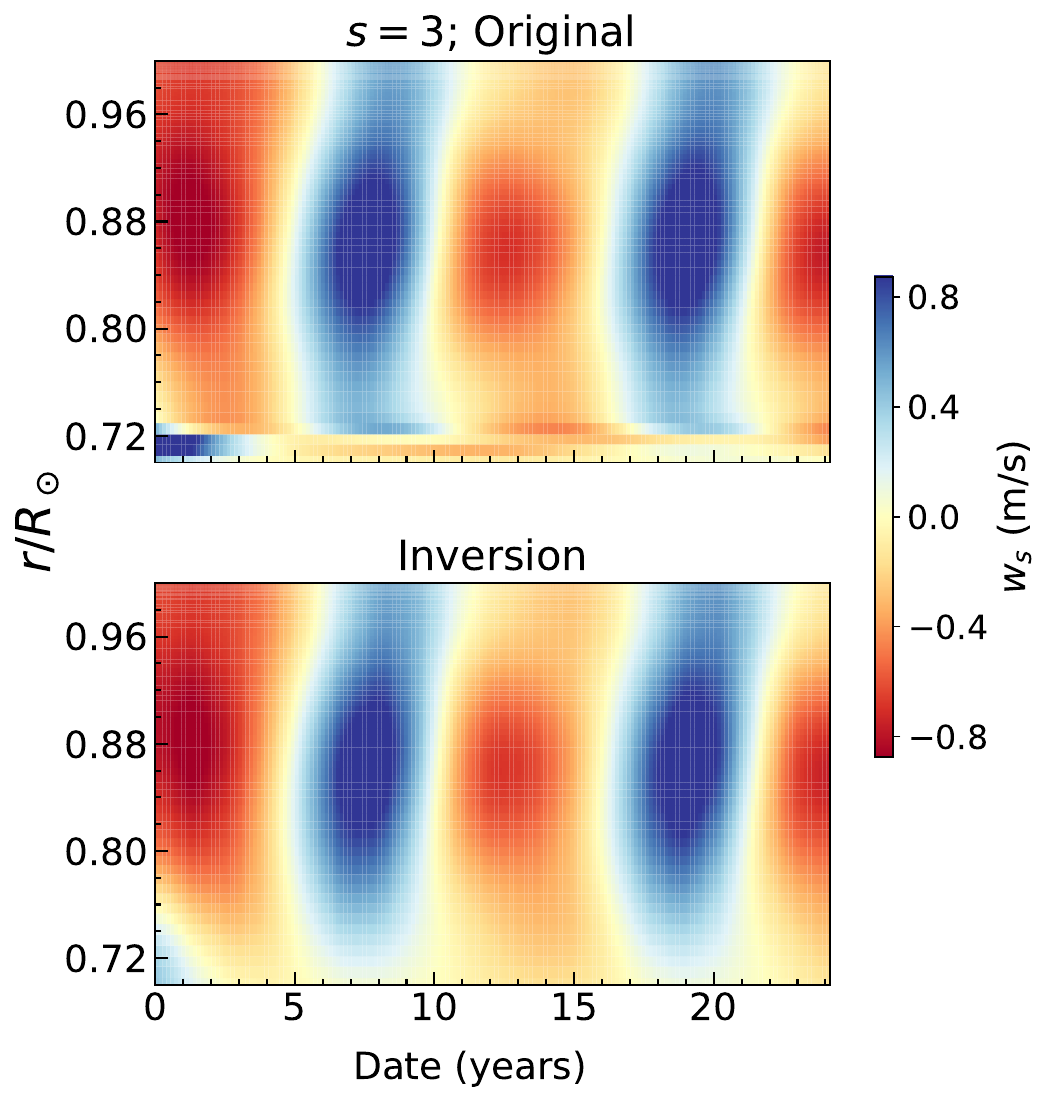}\includegraphics[scale=0.35]{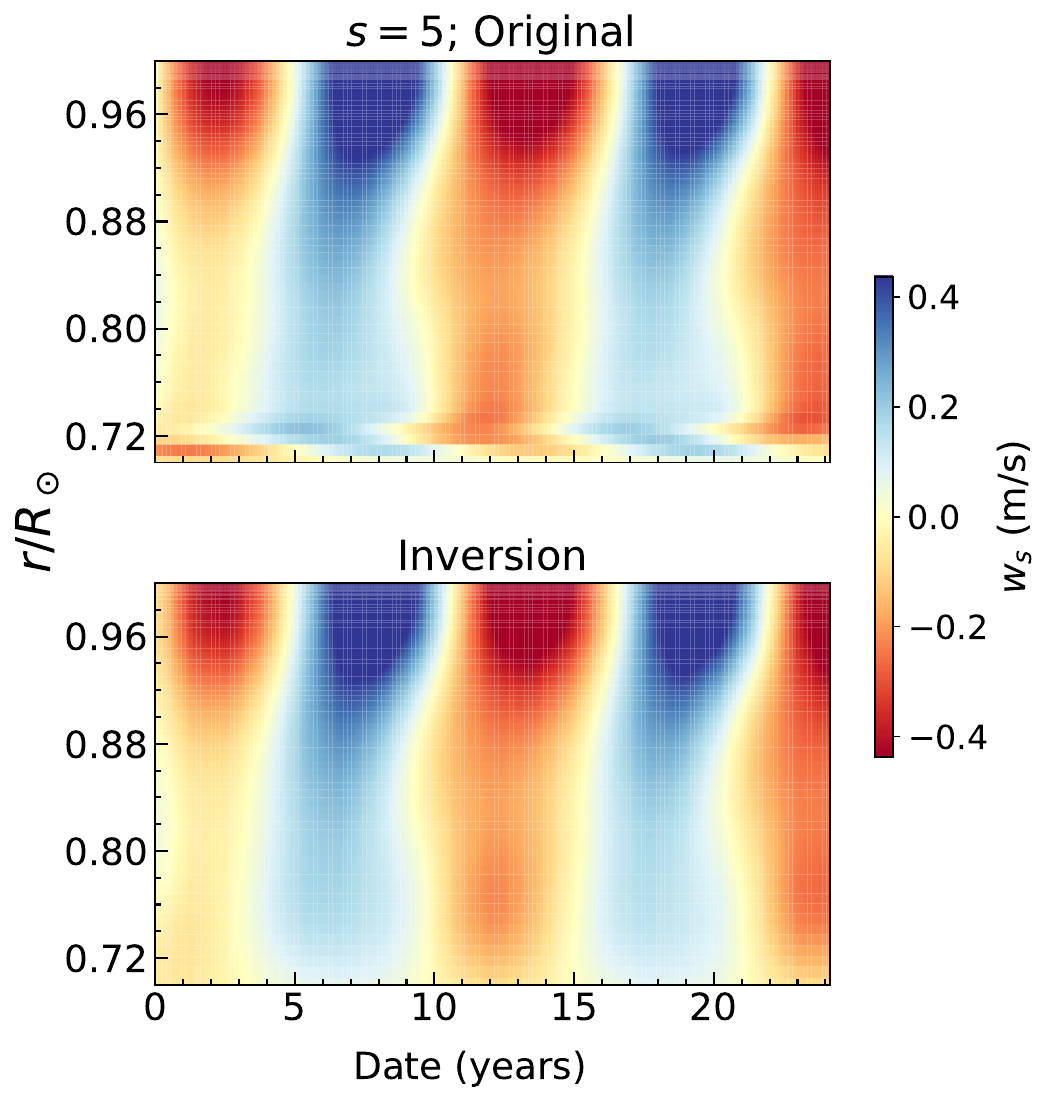}
    \caption{Zonal flow from the dynamo model undergoes spherical transformation according to Equation \ref{eq:v_w} to obtain $w_s$. We use these values to perform forward modeling and inversion. The inverted profile is then compared with the original profile in this Figure for harmonic degrees, $s=1$, $s=3$, and $s=5$. The upper panel shows the original profile, and the lower panel shows the reconstructed one.}
    \label{fig:wst_comp}
\end{figure}

Substituting the above equation into the first term of the misfit equation, we get
\begin{align*}
\chi^2 & =\sum_{n,\ell;t_{k}}\frac{(a_{s}(n,\ell;t_{k})-\sum_{i,j}c_{j,i}\int\mathcal{K}_{n,\ell}B_{j}(t_{k})B_{i}(r)r^{2}dr)^{2}}{\sigma_{s}(n,\ell;t_{k})^{2}}\\
\frac{\partial\chi}{\partial c_{j,i}} & =\sum_{n,\ell;t_{k}}\frac{(a_{s}(n,\ell;t_{k})-\sum_{i^{\prime},j^{\prime}}c_{j^{\prime},i^{\prime}}\int\mathcal{K}_{n,\ell}B_{j^{\prime}}(t_{k})B_{i^{\prime}}(r)r^{2}dr)}{\sigma_{s}(n,\ell,t_{k})^{2}}\int\mathcal{K}_{n,\ell}B_{j}(t_{k})B_{i}(r)r^{2}dr
\end{align*}
If we define matrix $A$, $D$  and $T$ as following 
\begin{align}
    A_{M,K} & =\int K_{M}B_{i}(r)B_{j}(t)r^{2}dr\\
    D_{ij} & =\int g(r) \frac{\partial^{2}B_{i}(r)}{\partial r^{2}}\frac{\partial^{2}B_{j}(r)}{\partial r^{2}} B_j(t) B_{j^\prime}(t)r^{2}dr\\
    T_{ij} & =\int f(r) \frac{\partial B_{i}(t)}{\partial t}\frac{\partial^{2}B_{j}(r)}{\partial r^{2}} B_j(t) B_{j^\prime}(t)r^{2}dr\\
    \label{eq:A_matrix}
\end{align}
where $M$ is a multi-index variable for $(n,\ell)$ and $K$ is a multi-index variable for $(i,j)$ and if we assume there are total $N_0$ such measurements, we can rewrite above equation as 
\begin{equation}
\sum_{i=1}^{N}\left[\sum_{M\in N_0}(A^{T})_{jM}\Lambda_{MM}^{-1}A_{Mi}+\lambda_r D_{ij} +\lambda_t T_{ij}\right]b_{i}=\sum_{M\in N_0}A_{jM}^{T}\Lambda_{MM}^{-1}a_{M},
\end{equation}
where $\Lambda$ is a diagonal matrix
\begin{equation}
\Lambda^{-1}=\begin{pmatrix}1/\sigma_{1}^{2} & 0 & 0 & 0& \dots\\
0 & 1/\sigma_{2}^{2} & 0 & 0 & \dots\\
0 & 0 & 1/\sigma_{3}^{2} & 0 & \dots\\
0 & 0 & 0 & 1/\sigma_{4}^{2} & \dots\\
\dots & \dots & \dots &\dots &\dots
\end{pmatrix}
\end{equation}
$\Lambda^{-1}$ is a $N_0\times N_0$ matrix. The above equation can be written in a matrix form as 
\begin{align}
\left[(A^{T}\Lambda^{-1}A)_{ji}+\lambda_r D_{ji}+\lambda_t T_{ji}\right]b_{i} & =(A^{T}\Lambda^{-1})_{jM} a_{M}\nonumber\\
\left[A^{T}\Lambda^{-1}A+\lambda_r D +\lambda_t T\right ]b & =A^{T}\Lambda^{-1}a
\label{eq:inv_app}
\end{align}
where matrix $b=(b_1,b_2,\dots,b_N)$. Solving these equations provides us with the values of $b$, which are used to reconstruct $w_{s}(r)$ from equation \ref{eq:v_w}. Assuming $f(r)=1$ and $g(r)=1$ in Equation \ref{eq:chi}, we perform an inversion using a known rotation profile. The results are shown in Figure \ref{fig:wst_comp}. We compare the time-varying radial profiles of $w_s$ for $s=1,3$ and $5$  between the inverted and the original profiles. The original profile is recovered with good accuracy.

 \section{Temporal Smoothing of Traditional Inversion Results} \label{sec:app_smooth}
Here, we present an alternative approach to smoothing the solution obtained from each time-segmented dataset, such as the inverted profile for each 72-day segment, which is generally considered before. In order to smooth the solution in time, Gaussian smoothing was used earlier. 
Since the results are generally somewhat noisy over time, we aim to derive a smoother solution from the inverted \( w_s(r,t) \) profiles. To achieve this, we represent the smooth solution using a cubic B-spline basis in both radius and time. Rewriting Equation \ref{eq:wst_app}, we obtain: 
\begin{equation} 
w_s(r,t) = \sum_i c_i \Phi_i, 
\end{equation} 
where
\begin{equation} 
\Phi_i(r,t) = B_j(t) B_k(r),
\end{equation} 
and  $i$ corresponds to a multi-index associated with (j,k).
We minimize the following misfit functions
\begin{align*}
\chi^{2} & =\sum_{r,t}\left(w_{s}(r,t)-\sum_i c_i \Phi_i(r,t)\right)^{2}+\lambda_{r}f(r)\sum_{t}\left(\frac{\partial ^{2}w_{s}}{ \partial t^{2}}\right)^{2}\\
\sum_{i} & M_{ij}c_{i}+\sum_{i}\lambda_{t}M_{ij}^{\text{Reg}}c_i=W_{j}\\
(M+\lambda_{t}M^{\text{Reg}})c & =W
\end{align*}
with respect to the unknown coefficients $c_{i}$. The second term in the above equation ensures a smooth solution in time. Smoothing in radius is not required, as it is already incorporated through the second-derivative regularization in the RLS inversion. The matrix $W$ in the right-hand side is
\begin{equation}
    W_{i}=\int_{0}^{T} \int_{0}^{R_\odot} w_{s}(r,t)\Phi_{i}(r,t) dt dr
\end{equation}

and the matrix $M$ is 
\begin{equation}
    M_{ij}=\int_{0}^{T} \int_{0}^{R} \Phi_{i}(t,r)\Phi_{j}(t,r) dt dr
\end{equation}
and $M_\text{Reg}$ is 
\begin{equation}
    M^{\text{Reg}}_{ij}= \int_{0}^{T}\int_{0}^{R_\odot} f(r)\frac{\partial^2\Phi_{i}}{\partial t^2}\frac{\partial^2\Phi_{j}}{\partial t^2}
\end{equation}
We adjust $\lambda_t$ to control the smoothness of the solution in time.

\section{Different Radial Function for Smoothing in Time}
\begin{figure}
    \centering
    \includegraphics[scale=0.45]{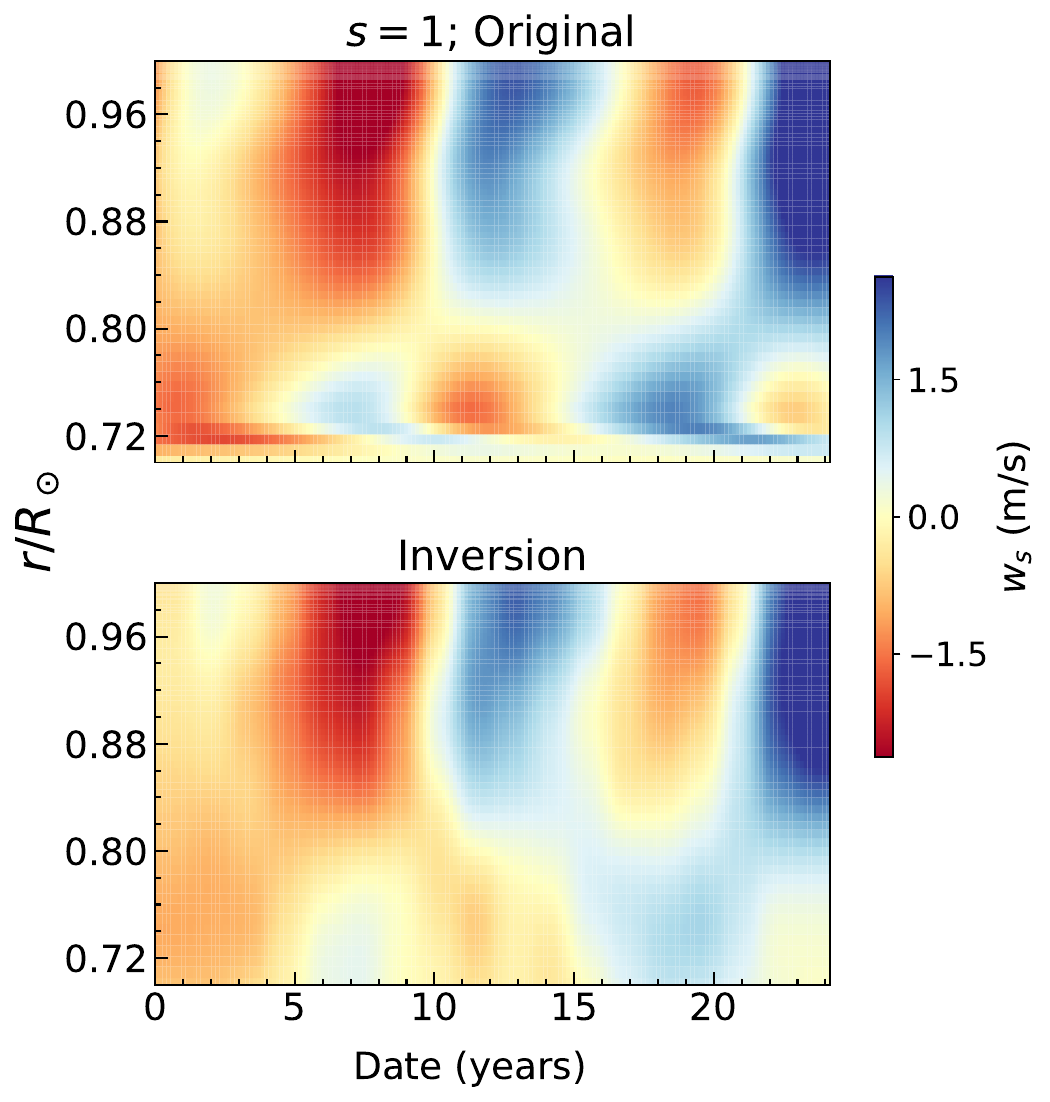}\includegraphics[scale=0.45]{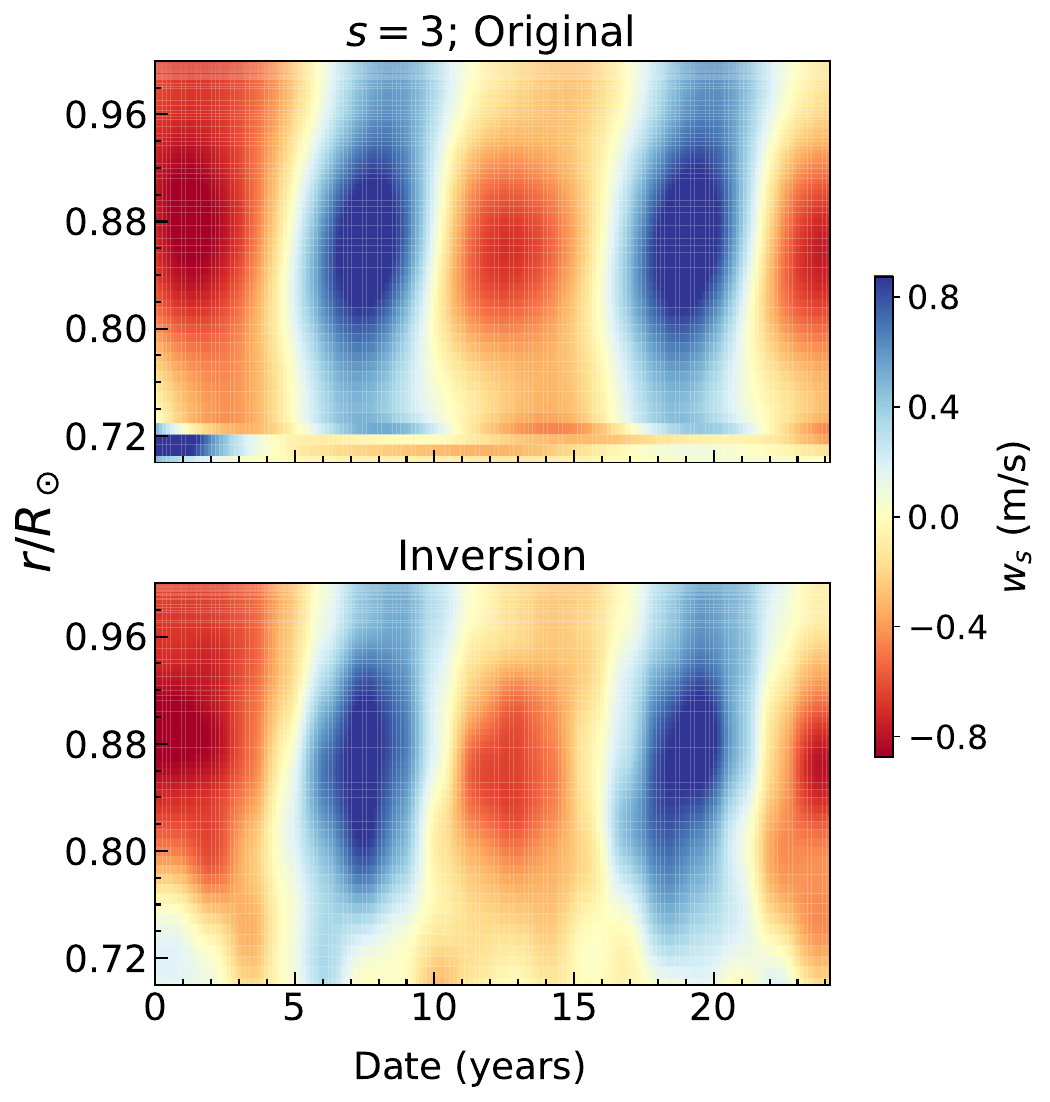}\\
    \includegraphics[scale=0.45]{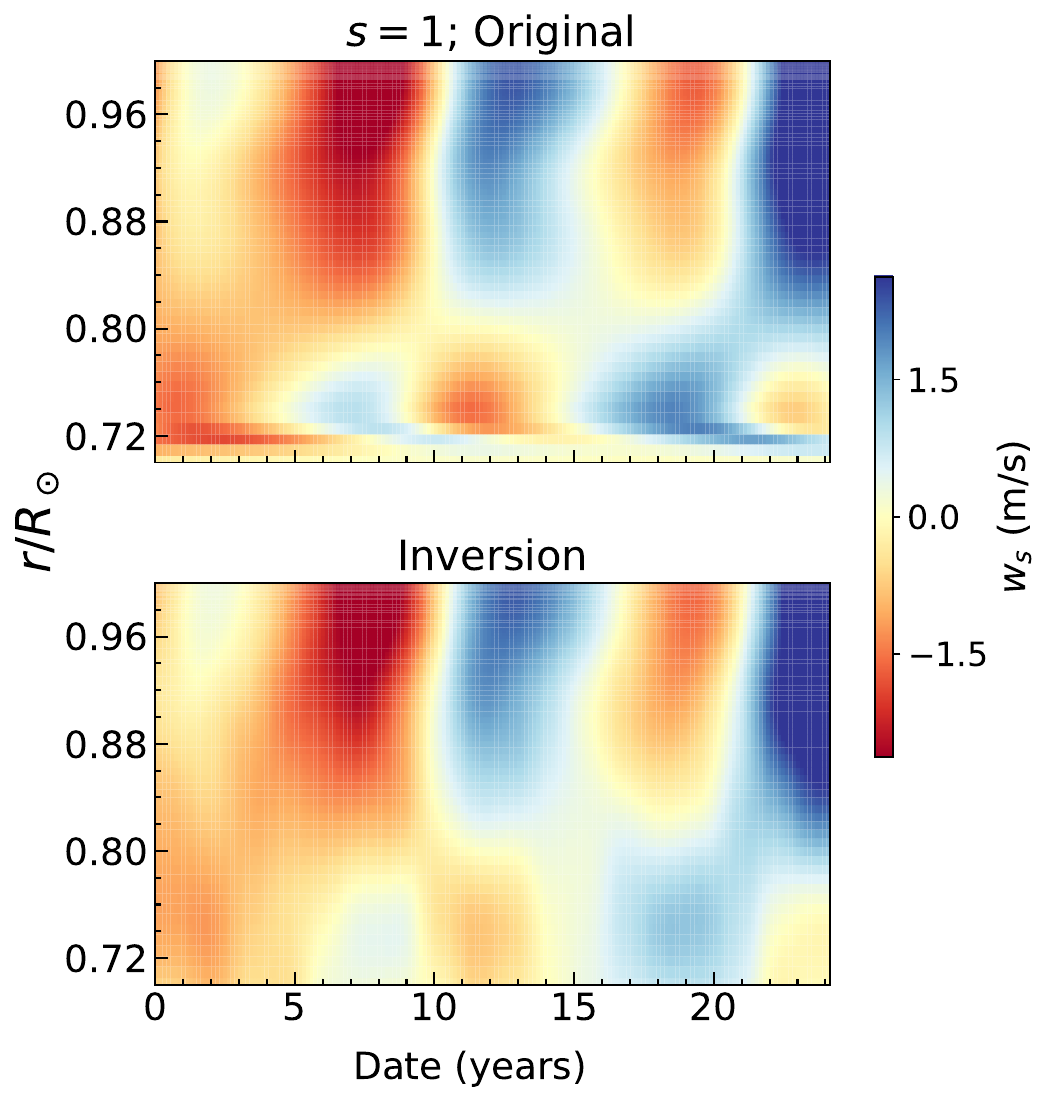}\includegraphics[scale=0.45]{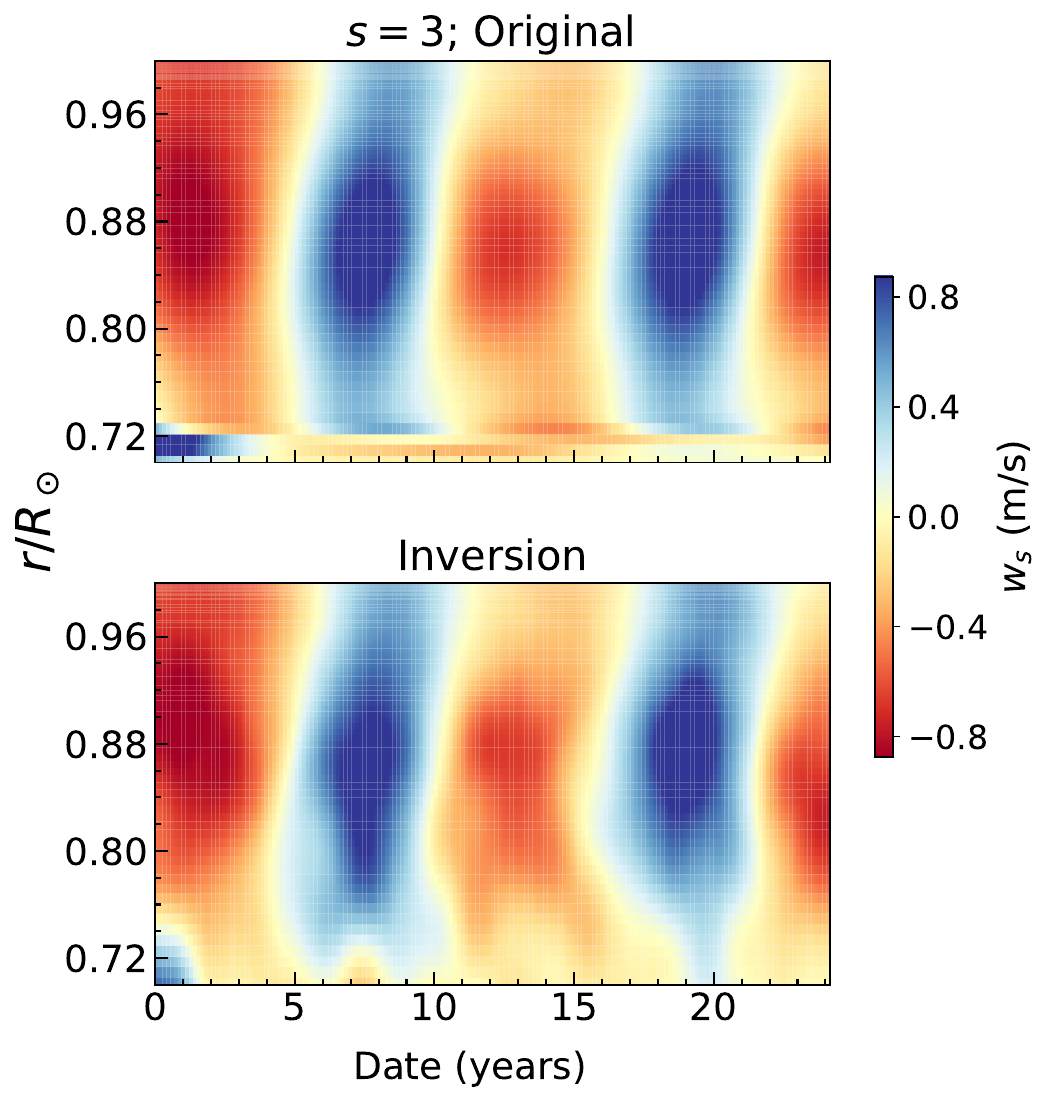}
    \caption{Inversion results using different functional forms of $f(r)$ in Equation~\ref{eq:chi} are shown: the top panel uses $f(r)=r^{-1}$, while the bottom panel uses $f(r)=r^{-2}$ for smoothing. We compare the inverted profiles with the original ones for harmonic degrees $s=1$ and $s=3$.}
    \label{fig:smooth_fr}
\end{figure}

We adopt different functional forms for $f(r)$ in Equation \ref{eq:chi} to achieve temporal smoothing. Specifically, we consider two functions: $r^{-1}$ and $r^{-2}$.The choice of these radial functions provides increased smoothing near the tachocline, as both functions increase with depth. This is beneficial in regions where fewer modes penetrate, resulting in limited information about deeper layers. Consequently, enhanced smoothing in this region ensures a more stable solution. Regarding the averaging kernels, their width is greater near the tachocline than at the surface, as demonstrated in our previous work \citep[Figure 9 of][]{mandal24_dw}. Figure \ref{fig:smooth_fr} presents results obtained using different functional forms for $f(r)$. While the radial profile for $s=1$ is well captured in both cases, the $s=3$ profile is accurately recovered when using $f(r) = r^{-1}$ but not as well with  $f(r) = r^{-2}$. Comparing the results without a functional form (Figure \ref{fig:wst_comp}) and with a functional form (Figure \ref{fig:smooth_fr}), we find that applying smoothing makes it impossible to capture the sharp transition near the tachocline, inevitably leading to a smooth solution in that region.

% \begin{figure}
%     \centering
%     \includegraphics[scale=0.45]{fig/wst_reg_s1.pdf}
    
%     \caption{We vary $\lambda_t$ to examine its effect on our inverted profile. A higher $\lambda_t$ results in a smoother inverted profile compared to a lower $\lambda_t$ as expected.}
%     \label{fig:time_smooth}
% \end{figure}
 \bibliography{references}

\begin{thebibliography}{}
\expandafter\ifx\csname natexlab\endcsname\relax\def\natexlab#1{#1}\fi
\providecommand{\url}[1]{\href{#1}{#1}}
\providecommand{\dodoi}[1]{doi:~\href{http://doi.org/#1}{\nolinkurl{#1}}}
\providecommand{\doeprint}[1]{\href{http://ascl.net/#1}{\nolinkurl{http://ascl.net/#1}}}
\providecommand{\doarXiv}[1]{\href{https://arxiv.org/abs/#1}{\nolinkurl{https://arxiv.org/abs/#1}}}

\bibitem[{{Antia} \& {Basu}(2022)}]{antia2022}
{Antia}, H.~M., \& {Basu}, S. 2022, \apj, 924, 19, \dodoi{10.3847/1538-4357/ac32c3}

\bibitem[{{Antia} {et~al.}(2008){Antia}, {Basu}, \& {Chitre}}]{antia08}
{Antia}, H.~M., {Basu}, S., \& {Chitre}, S.~M. 2008, \apj, 681, 680, \dodoi{10.1086/588523}

\bibitem[{{Babcock}(1961)}]{babcock61}
{Babcock}, H.~W. 1961, \apj, 133, 572, \dodoi{10.1086/147060}

\bibitem[{{Backus} \& {Gilbert}(1968)}]{backus_1968_ola}
{Backus}, G., \& {Gilbert}, F. 1968, Geophysical Journal, 16, 169, \dodoi{10.1111/j.1365-246X.1968.tb00216.x}

\bibitem[{{Backus} \& {Gilbert}(1970)}]{backus_1970_OLA}
---. 1970, Philosophical Transactions of the Royal Society of London Series A, 266, 123, \dodoi{10.1098/rsta.1970.0005}

\bibitem[{{Barekat} {et~al.}(2014){Barekat}, {Schou}, \& {Gizon}}]{barekat2014}
{Barekat}, A., {Schou}, J., \& {Gizon}, L. 2014, \aap, 570, L12, \dodoi{10.1051/0004-6361/201424839}

\bibitem[{{Basu}(2016)}]{basu_2016}
{Basu}, S. 2016, Living Reviews in Solar Physics, 13, 2, \dodoi{10.1007/s41116-016-0003-4}

\bibitem[{{Basu} \& {Antia}(2019)}]{basu19}
{Basu}, S., \& {Antia}, H.~M. 2019, \apj, 883, 93, \dodoi{10.3847/1538-4357/ab3b57}

\bibitem[{{Brandenburg}(2005)}]{brandenburg_2005}
{Brandenburg}, A. 2005, \apj, 625, 539, \dodoi{10.1086/429584}

\bibitem[{{Charbonneau} \& {Barlet}(2011)}]{paul11}
{Charbonneau}, P., \& {Barlet}, G. 2011, Journal of Atmospheric and Solar-Terrestrial Physics, 73, 198, \dodoi{10.1016/j.jastp.2009.12.020}

\bibitem[{{Choudhuri} \& {Dikpati}(1999)}]{arnab_dikpati99}
{Choudhuri}, A.~R., \& {Dikpati}, M. 1999, \solphys, 184, 61, \dodoi{10.1023/A:1005092601436}

\bibitem[{{Corbard} \& {Thompson}(2002)}]{corbard2002}
{Corbard}, T., \& {Thompson}, M.~J. 2002, \solphys, 205, 211, \dodoi{10.1023/A:1014224523374}

\bibitem[{{Dikpati} {et~al.}(2002){Dikpati}, {Corbard}, {Thompson}, \& {Gilman}}]{Dikpati2002}
{Dikpati}, M., {Corbard}, T., {Thompson}, M.~J., \& {Gilman}, P.~A. 2002, \apjl, 575, L41, \dodoi{10.1086/342555}

\bibitem[{{Dikpati} {et~al.}(2009){Dikpati}, {Gilman}, {Cally}, \& {Miesch}}]{dikpati09}
{Dikpati}, M., {Gilman}, P.~A., {Cally}, P.~S., \& {Miesch}, M.~S. 2009, \apj, 692, 1421, \dodoi{10.1088/0004-637X/692/2/1421}

\bibitem[{{Guerrero} {et~al.}(2016){Guerrero}, {Smolarkiewicz}, {de Gouveia Dal Pino}, {Kosovichev}, \& {Mansour}}]{gustavo16}
{Guerrero}, G., {Smolarkiewicz}, P.~K., {de Gouveia Dal Pino}, E.~M., {Kosovichev}, A.~G., \& {Mansour}, N.~N. 2016, \apjl, 828, L3, \dodoi{10.3847/2041-8205/828/1/L3}

\bibitem[{{Hill} {et~al.}(1996){Hill}, {Stark}, {Stebbins}, {Anderson}, {Antia}, {Brown}, {Duvall}, {Haber}, {Harvey}, {Hathaway}, {Howe}, {Hubbard}, {Jones}, {Kennedy}, {Korzennik}, {Kosovichev}, {Leibacher}, {Libbrecht}, {Pintar}, {Rhodes}, {Schou}, {Thompson}, {Tomczyk}, {Toner}, {Toussaint}, \& {Williams}}]{hill_1996_GONG}
{Hill}, F., {Stark}, P.~B., {Stebbins}, R.~T., {et~al.} 1996, Science, 272, 1292, \dodoi{10.1126/science.272.5266.1292}

\bibitem[{{Howard} \& {Labonte}(1980)}]{howard80}
{Howard}, R., \& {Labonte}, B.~J. 1980, \apjl, 239, L33, \dodoi{10.1086/183286}

\bibitem[{{Howe} {et~al.}(2018){Howe}, {Hill}, {Komm}, {Chaplin}, {Elsworth}, {Davies}, {Schou}, \& {Thompson}}]{howe18}
{Howe}, R., {Hill}, F., {Komm}, R., {et~al.} 2018, \apjl, 862, L5, \dodoi{10.3847/2041-8213/aad1ed}

\bibitem[{{Jha} \& {Choudhuri}(2021)}]{jha_2021}
{Jha}, B.~K., \& {Choudhuri}, A.~R. 2021, \mnras, 506, 2189, \dodoi{10.1093/mnras/stab1717}

\bibitem[{{Karak} \& {Cameron}(2016)}]{karak2016}
{Karak}, B.~B., \& {Cameron}, R. 2016, \apj, 832, 94, \dodoi{10.3847/0004-637X/832/1/94}

\bibitem[{{Kitchatinov}(2016)}]{kitchatinov_2016}
{Kitchatinov}, L.~L. 2016, Astronomy Letters, 42, 339, \dodoi{10.1134/S1063773716050054}

\bibitem[{{Kitiashvili} {et~al.}(2023){Kitiashvili}, {Kosovichev}, {Wray}, {Sadykov}, \& {Guerrero}}]{irina2023}
{Kitiashvili}, I.~N., {Kosovichev}, A.~G., {Wray}, A.~A., {Sadykov}, V.~M., \& {Guerrero}, G. 2023, \mnras, 518, 504, \dodoi{10.1093/mnras/stac2946}

\bibitem[{{Komm}(2023)}]{komm2023}
{Komm}, R. 2023, \solphys, 298, 119, \dodoi{10.1007/s11207-023-02213-7}

\bibitem[{{Korzennik}(2023)}]{sylvain23}
{Korzennik}, S.~G. 2023, Frontiers in Astronomy and Space Sciences, 9, 1031313, \dodoi{10.3389/fspas.2022.1031313}

\bibitem[{{Korzennik} \& {Eff-Darwich}(2024)}]{sylvain_2024_inv}
{Korzennik}, S.~G., \& {Eff-Darwich}, A. 2024, \solphys, 299, 86, \dodoi{10.1007/s11207-024-02334-7}

\bibitem[{{Korzennik} {et~al.}(2013){Korzennik}, {Rabello-Soares}, {Schou}, \& {Larson}}]{korzennik2013}
{Korzennik}, S.~G., {Rabello-Soares}, M.~C., {Schou}, J., \& {Larson}, T.~P. 2013, \apj, 772, 87, \dodoi{10.1088/0004-637X/772/2/87}

\bibitem[{{Kosovichev} \& {Pipin}(2019)}]{sasha2019}
{Kosovichev}, A.~G., \& {Pipin}, V.~V. 2019, \apjl, 871, L20, \dodoi{10.3847/2041-8213/aafe82}

\bibitem[{{Kosovichev} \& {Schou}(1997)}]{sasha97}
{Kosovichev}, A.~G., \& {Schou}, J. 1997, \apjl, 482, L207, \dodoi{10.1086/310708}

\bibitem[{{Larson} \& {Schou}(2015)}]{larson2015}
{Larson}, T.~P., \& {Schou}, J. 2015, \solphys, 290, 3221, \dodoi{10.1007/s11207-015-0792-y}

\bibitem[{{Larson} \& {Schou}(2018)}]{larson18}
---. 2018, \solphys, 293, 29, \dodoi{10.1007/s11207-017-1201-5}

\bibitem[{{Mandal} {et~al.}(2024){Mandal}, {Kosovichev}, \& {Pipin}}]{mandal24_dw}
{Mandal}, K., {Kosovichev}, A.~G., \& {Pipin}, V.~V. 2024, \apj, 973, 36, \dodoi{10.3847/1538-4357/ad5f2c}

\bibitem[{{Matilsky} {et~al.}(2019){Matilsky}, {Hindman}, \& {Toomre}}]{matilsky_2019}
{Matilsky}, L.~I., {Hindman}, B.~W., \& {Toomre}, J. 2019, \apj, 871, 217, \dodoi{10.3847/1538-4357/aaf647}

\bibitem[{{Parker}(1955)}]{parker_55_dynamo}
{Parker}, E.~N. 1955, \apj, 122, 293, \dodoi{10.1086/146087}

\bibitem[{{Pijpers} \& {Thompson}(1994)}]{pijpers_1994_SOLA}
{Pijpers}, F.~P., \& {Thompson}, M.~J. 1994, \aap, 281, 231

\bibitem[{{Pipin} \& {Kosovichev}(2011)}]{pipin2011}
{Pipin}, V.~V., \& {Kosovichev}, A.~G. 2011, \apjl, 727, L45, \dodoi{10.1088/2041-8205/727/2/L45}

\bibitem[{{Pipin} \& {Kosovichev}(2019)}]{pipin2019}
---. 2019, \apj, 887, 215, \dodoi{10.3847/1538-4357/ab5952}

\bibitem[{{Pipin} \& {Kosovichev}(2020)}]{pipin20}
---. 2020, \apj, 900, 26, \dodoi{10.3847/1538-4357/aba4ad}

\bibitem[{{Rabello Soares} {et~al.}(2024){Rabello Soares}, {Basu}, \& {Bogart}}]{cristina2024}
{Rabello Soares}, M.~C., {Basu}, S., \& {Bogart}, R.~S. 2024, \apj, 967, 143, \dodoi{10.3847/1538-4357/ad3d59}

\bibitem[{{Ritzwoller} \& {Lavely}(1991)}]{ritzwoller91}
{Ritzwoller}, M.~H., \& {Lavely}, E.~M. 1991, \apj, 369, 557, \dodoi{10.1086/169785}

\bibitem[{{Rozelot} {et~al.}(2025){Rozelot}, {Kosovichev}, \& {Kitiashvili}}]{rozelot2025}
{Rozelot}, J.-P., {Kosovichev}, A., \& {Kitiashvili}, I. 2025, arXiv e-prints, arXiv:2501.08021, \dodoi{10.48550/arXiv.2501.08021}

\bibitem[{{Schatten} {et~al.}(1978){Schatten}, {Scherrer}, {Svalgaard}, \& {Wilcox}}]{schatten78}
{Schatten}, K.~H., {Scherrer}, P.~H., {Svalgaard}, L., \& {Wilcox}, J.~M. 1978, \grl, 5, 411, \dodoi{10.1029/GL005i005p00411}

\bibitem[{{Scherrer} {et~al.}(1995){Scherrer}, {Bogart}, {Bush}, {Hoeksema}, {Kosovichev}, {Schou}, {Rosenberg}, {Springer}, {Tarbell}, {Title}, {Wolfson}, {Zayer}, \& {MDI Engineering Team}}]{scherrer95}
{Scherrer}, P.~H., {Bogart}, R.~S., {Bush}, R.~I., {et~al.} 1995, \solphys, 162, 129, \dodoi{10.1007/BF00733429}

\bibitem[{{Scherrer} {et~al.}(2012){Scherrer}, {Schou}, {Bush}, {Kosovichev}, {Bogart}, {Hoeksema}, {Liu}, {Duvall}, {Zhao}, {Title}, {Schrijver}, {Tarbell}, \& {Tomczyk}}]{scherrer2012}
{Scherrer}, P.~H., {Schou}, J., {Bush}, R.~I., {et~al.} 2012, \solphys, 275, 207, \dodoi{10.1007/s11207-011-9834-2}

\bibitem[{{Schou} {et~al.}(1998){Schou}, {Antia}, {Basu}, {Bogart}, {Bush}, {Chitre}, {Christensen-Dalsgaard}, {Di Mauro}, {Dziembowski}, {Eff-Darwich}, {Gough}, {Haber}, {Hoeksema}, {Howe}, {Korzennik}, {Kosovichev}, {Larsen}, {Pijpers}, {Scherrer}, {Sekii}, {Tarbell}, {Title}, {Thompson}, \& {Toomre}}]{schou98}
{Schou}, J., {Antia}, H.~M., {Basu}, S., {et~al.} 1998, \apj, 505, 390, \dodoi{10.1086/306146}

\bibitem[{{Schou} {et~al.}(2012){Schou}, {Scherrer}, {Bush}, {Wachter}, {Couvidat}, {Rabello-Soares}, {Bogart}, {Hoeksema}, {Liu}, {Duvall}, {Akin}, {Allard}, {Miles}, {Rairden}, {Shine}, {Tarbell}, {Title}, {Wolfson}, {Elmore}, {Norton}, \& {Tomczyk}}]{hmi}
{Schou}, J., {Scherrer}, P.~H., {Bush}, R.~I., {et~al.} 2012, \solphys, 275, 229, \dodoi{10.1007/s11207-011-9842-2}

\bibitem[{{Vorontsov} {et~al.}(2002){Vorontsov}, {Christensen-Dalsgaard}, {Schou}, {Strakhov}, \& {Thompson}}]{vorontsov02}
{Vorontsov}, S.~V., {Christensen-Dalsgaard}, J., {Schou}, J., {Strakhov}, V.~N., \& {Thompson}, M.~J. 2002, Science, 296, 101, \dodoi{10.1126/science.1069190}

\end{thebibliography}
  \end{document}